\documentclass[12pt]{article}
\usepackage{epsfig}
\usepackage{graphicx,floatflt,amssymb,epsf}
\usepackage{subfigure}
\usepackage{graphics}
\global\parskip 6pt

\normalsize

\def\br{\begin{thebibliography}{abc}}
\def\er{\end{thebibliography}}

\def\be{\begin{equation}}

\def\ee{\end{equation}}

\def\bea{\begin{eqnarray}}

\def\eea{\end{eqnarray}}
\def\tg{\tilde\gamma}
\def\tn{\tilde\nabla}
\def\p{\tilde\psi}
\def\t{\tilde}
\begin{document}
\begin{flushright}
\hfill{SINP-TNP/08-17}\\
\end{flushright}
\vspace*{1cm}
\thispagestyle{empty}
\centerline{\large \bf A comparative study of Dirac quasinormal modes of }
\centerline{\large\bf charged black holes in higher dimensions}
\bigskip
\begin{center}
Sayan K. Chakrabarti\footnote{Email: sayan.chakrabarti@saha.ac.in}\\
\vspace*{.5cm}
{\em Theory Division\\
Saha Institute of Nuclear Physics\\
1/AF Bidhannagar, Calcutta - 700064, India}\\
\vspace*{.5cm}
\end{center}
\vskip.5cm

\begin{abstract}
In this work we study the Dirac quasinormal modes of higher dimensional charged black holes. Higher dimensional Reissner-Nordstr\"{o}m type black holes as well as charged black holes in Einstein Gauss-Bonnet theories are studied for Fermionic perturbations using WKB methods. A comparative study of the quasinormal modes in the two different theories of gravity has been performed. The beahviour of the frequencies with the variation of black hole parameters as well as with the variation of spacetime dimensions are done. We also study the large multipole number limit of the black hole potential in order to look for an analytic expression for the frequencies.    
\end{abstract}
\vspace*{.3cm}
\begin{center}
%Keywords: 
\end{center}
\vspace*{.3cm}
\newpage

\section{Introduction} 

The dynamical evolution of perturbation in a black hole background \cite{kk, nol}
mainly consists of three stages - the first, an initial wave
burst which depends on the original form of the field perturbation,
second, the damped oscillation whose frequency and damping
depends entirely on the background spacetime and not on the perturbing
field. The final stage involves a power law tail behaviour of the waves at very
late times. It is the second stage which is of most importance in
studying black hole physics. This is characterized by a discrete set
of frequencies termed quasinormal frequencies. On perturbing the black
hole background, either through external fields or via metric
perturbation, the black hole system responds by emitting damped
oscillations \cite{kk,nol}. The frequencies consists of a real part
which represents the frequency of the oscillations and an imaginary
part representing the damping.  For a large class of black holes, the
equations governing the perturbations can be cast into the
Schr\"{o}dinger like wave equation.  For asymptotically flat
space-times, the quasinormal modes (QNMs) are solutions of the
corresponding wave equation with complex frequencies which are purely
ingoing at the horizon and purely outgoing at spatial infinity
\cite{rg,vish}. One of the most important interest in studying
QNMs arises from the fact that there is a possibility of detecting
them in gravitational wave detectors \cite{sterio, fera}. Apart
from that, QNM's have been found to be a useful probe of the
underlying space-time geometry. QN frequencies carry unique
information about black hole parameters \cite{card_thes}. It has also been shown that quasinormal modes in Anti-de Sitter
space-time appear naturally in the description of the corresponding
dual conformal field theories living on the boundary \cite{danny,
  danny1}. Therefore, many researchers found it interesting to study
QNMs in the background of asymptotically AdS black holes
\cite{horo, card1}. The QNMs of black hole in
asymptotically de Sitter spaces were also studied accordingly
\cite{kono3}, since there
were evidences that our universe is governed by positive cosmological
constant \cite{perl}. Let us also mention here for the sake of
completeness that in spite of their classical origin, it has been
suggested that QNM's might provide a glimpse into the quantum nature
of the black hole \cite{hod}.

Black hole QNMs have been studied since long time. A lot of
studies were made for perturbations of black hole background with
fields which are of integer spins (scalar, electromagnetic and
gravitational). Compared to this, the study of Dirac QNMs in black
hole backgrounds is rather very limited. The Dirac QNM for (2+1)
dimensional BTZ black hole in AdS space was studied both numerically
and analytically in \cite{danny, card1}. Studying Dirac QNMs in four
dimensions started with the work of Cho \cite{cho} who first studied
massive and massless Dirac QNMs in a Schwarzschild black hole
background using the WKB approach \cite{will,will2} followed by Jing
\cite{jing}, who modified the approach using continued fraction
\cite{lever} and Hill determinant method \cite{hill}. For the massless
Dirac field, it was shown that the real part of the QN frequency
increases with the multipole number $l$ while the imaginary part
increases with the overtone numbers $n$. For the massive Dirac field
the real part of the QN frequency increases with the mass of the
field, while the damping decreases, which implies that it would be
possible to detect QN frequencies due to perturbations of black hole
background with massive Dirac fields, since fields with higher masses
will decay slowly. The high overtones of Dirac perturbations of a
Schwarzschild black hole were studied in \cite{khc}. Then there were
lots of similar works on the Dirac perturbations of four dimensional
black holes both in asymptotically flat and non-flat black hole
backgrounds \cite{wu,jing3}. 

All the abovementioned studies of Dirac QNMs were done in the four
dimensional black hole backgrounds while, the study of Dirac
perturbations in higher dimensional black hole backgrounds are even more
limited. String theory requires the existence of extra spatial
dimensions and for a long time it was thought that the only possible
way to think of these as extra spatial dimensions tightly curled with
the radius of curvature around string scale. However later it was
realized that these extra spatial dimensions need not be of the order
of string scale but they can be as large as a few millimeters \cite{antoniadis1,led,antoniadis}.
% The
%theory with large extra dimensions \cite{led} predicts the existence
%of $d$ extra spacelike dimensions. According to these theories the
%standard model fields are constrained to live on a four dimensional
%hypersurface called a brane which is embedded in a higher dimensional
%bulk, while the gravitons are free to propagate on the brane as well
%as on the bulk. 
An important feature of the large extra dimensions is that it implies that the fundamental Planck
scale is much lower than the four dimensional Planck scale, in fact it
might be of the order of a few TeV. This provides a tool for looking into
stringy effects which might become observable at the LHC. One of the most
interesting effects will be the production of microscopic black holes
at the LHC. Since gravity will be sensitive to macroscopic extra
dimensions, these black holes produced at the colliders will
essentially be higher dimensional. The study of QNMs of these kind of
black holes projected on the four dimensional brane with the
perturbations due to brane localized standard model fields were made
in \cite{kanti1, kanti2, zhidGB}. It was shown that the increase in number of
extra spatial hidden dimensions dampens the QN frequencies produced
via perturbations of all kinds of brane localized standard model
fields. The presence of charge $Q$ in the black hole background also
affects the QN spectrum and it is significantly different from the
behavior of charged black holes in four dimensions. 

However, the study of Dirac QNMs of purely higher dimensional black
holes were not present in the literature until the work of Cho et al
\cite{split}. In that work they have studied QNMs of Schwarzschild
black holes using the conformal properties of the spinor field. Such an
idea is perfectly general and in principle can be applied to all
higher dimensional spherically symmetric black holes. In this paper we
will use the abovementioned idea to study QNMs of
higher dimensional charged black holes arising out of Einstein Hilbert
action as well as from higher derivative corrections to such actions.
In particular we will study the QNMs of higher dimensional Reissner-
N\"{o}rdtrom type black holes \cite{mype} and the charged Gauss-Bonnet black holes \cite{zwei,wheeler,wheeler1,wiltsh}. The motivation of the paper is three fold, namely 

To study the Dirac QNM in charged black hole backgrounds in higher dimensions. As we have mentioned that study of Dirac perturbations in higher dimension is rather limited, this paper will try to fill in a gap in the literature by studying Dirac QNMs in purely higher dimensional charged black hole backgrounds in the framework of general relativity and its higher derivative corrected scenario.

To compare the results of charged black hole QNMs in two different scenarios, namely the black holes arising out of Einstein Hilbert action and in the higher derivative gravities, precisely the Gauss- Bonnet black holes due to Dirac perturbations. We also study the behaviour of the QN frequency with the Gauss-Bonnet coupling $\alpha$ for the charged Gauss-Bonnet black hole.
 
To compare our results with the available results for brane localized black holes studied by Kanti etal \cite{kanti1, kanti2} and Zhidenko \cite{zhidGB}.

The plan of the paper is as follows: in the next section we briefly discuss the Reissner Nordstr\"{o}m type solutions and charged Gauss Bonnet solution in dimensions $d>4$. In section 3 we present a brief discussion of WKB method along with a comparative study of the QNMs of Reissner-Nordstr\"{o}m and charged Gauss-Bonnet
black holes. Section 4 deals with the quasinormal modes in large multipole number limit and in section 5 we conclude the paper with a brief discussion on future directions. Finally in the appendix we give a brief review of the Dirac equations in higher dimensional curved background following \cite{split}.

\section{Charged Black Holes in Higher Dimensions}

In this section we  will discuss the charged black holes in higher dimensions. As we have mentioned that we will study charged black holes arising out of Einstein's theory of general relativity and those arising out of Einstein Gauss Bonnet theory. Let us first discuss the Reissner Nordstr\"{o}m type solutions in higher dimensional Einstein gravity. 

\subsection{Reissner-Nordstr\"{o}m type solutions}

One of the main goals of theoretical physics is to find out a theory which unifies gravity with all other fundamental forces in nature. String theory is one of the candidates for such a unified theory and it predicts that we live in a world which has dimensions greater than $3+1$. In this context, study of black hole physics is an important area. In this section we discuss higher dimensional black holes which are static and spherically symmetric. The study of such black holes began with the work of Tangherlini \cite{tnln} and later by Myers and Perry \cite{mype}.

Let us start with the static spherically symmetric metric 
\bea
ds^2=-f(r)dt^2+f^{-1}(r)dr^2+r^2d\bar\Omega_{d-2}^2, \label{methighd}
\eea
The vacuum Einstein equation then implies 
\bea
f=\left(1-\frac{2\mu}{r^{d-3}}\right),
\eea
provided that $d\ge 4$. The parameter $\mu$ is a constant of integration and is related to the mass $M$ of the black hole
\bea
M=\frac{(d-2)\mathcal{A}_{d-2}}{8\pi G_{d}}\mu,
\eea 
where $\mathcal{A}_{d-2}$ is the area of the unit $(d-2)$-sphere given by 
\bea
\mathcal{A}_{d-2}=\frac{2\pi^{(d-1)/2}}{\Gamma(\frac{d-1}{2})}
\eea
One can find the analog of Reissner-Nordstr\"{o}m solutions in higher dimensions also and there the metric is given by \cite{mype}
\bea
f=\left(1-\frac{2\mu}{r^{d-3}}+\frac{\theta^2}{r^{2(d-3)}}\right)
\eea
The electric charge of the black hole is given by
\bea
Q^2=\frac{(d-2)(d-3)}{8\pi G_{d}}\theta^2
\eea
For $\theta^2<\mu^2$, there is an outermost horizon situated at
\bea
r^{d-3}=\mu+(\mu^2-\theta^2)^{1/2} \label{extml}
\eea
This will be of importance for our analysis since we will be interested in the potential just outside the outermost horizon of the black hole. In our analysis of the QNM for the charged black hole, we will also have to keep in mind the relation (\ref{extml}), since this relation gives a constraint on the mass and charge of the hole. In other words, one can not work with arbitrary mass and charge parameter while working with the Reissner-Nordstr\"{o}m type metric. The non-extremality condition should always be maintained while working with such metrics.

\subsection{Charged Gauss-Bonnet black hole}

We will now briefly discuss the black holes which arise out of Einstein Gauss-Bonnet gravity. 
In space-time dimensions $d \geq 5$ the Einstein-Gauss-Bonnet action is given by
\be 
I=\frac{1}{16\pi G_d} \int d^dx \sqrt{-g} R + \alpha^{\prime}\int
  d^dx
  \sqrt{-g}(R_{\mu\nu\beta\gamma}R^{\mu\nu\beta\gamma}-4R_{\beta\gamma}R^{\beta\gamma}+R^2)  
,\label{alpha} 
\ee
where $G_d$ is the $d$-dimensional Newton's constant and the parameter
$\alpha^{\prime}$ denotes the Gauss-Bonnet coupling. We will choose
$G_d=1$ from now on and will consider only positive $\alpha^{\prime}$
which is consistent with the string expansion \cite{deser}.

The metric for spherically symmetric asymptotically flat Gauss-Bonnet
black hole solution of mass $M$ is given by Eqn. (\ref{methighd}), where
$f(r)$ has the form \cite{deser}
\be
f(r) = 1 + \frac{r^2}{2\alpha} -
\frac{r^2}{2\alpha}\sqrt{1+\frac{8\alpha M}{r^{d-1}}}, \label{fr}
\ee
where, 
\be
\alpha=16\pi G_d (d-3)(d-4)\alpha'.
\ee
For $\alpha'>0$, this black hole admits only a single horizon
\cite{deser}. The horizon $r=r_h$ is determined by the real positive
solution of the equation
\be
r_h^{d-3}+\alpha r_h^{d-5}=2M. \label{horizon}
\ee
The charged Gauss-Bonnet black hole has the following form of $f(r)$
\cite{wiltsh}: 
\be
f(r)=1+\frac{r^2}{2\alpha}-\frac{r^2}{2\alpha}\sqrt{1+\frac{8\alpha
    M}{r^{d-1}}-\frac{4\alpha Q^2}{2\pi (d-2)(d-3)r^{2d-4}}},
\ee
if $M>0$ and $\alpha >0$, then there will be a timelike singularity
which will be shielded by two horizons if $Q<Q_{ex}$. Here $Q_{ex}$ is
the extremal value of the charge determined from \cite{wiltsh}:
\be
r_{ex}^{2(d-3)}+\frac{d-5}{d-3}\alpha
r_{ex}^{2(d-4)}-\frac{Q_{ex}^2}{2\pi (d-2)(d-3)}=0,
\ee
where,
\be
r_{ex}^{d-3}=-\frac{1}{2}(d-5)M+\left[\frac{1}{4}(d-5)^2M^2+\frac{(d-4)Q_{ex}^2}{2\pi(d-2)(d-3)}\right]^{1/2}.
\ee
Again, this equation will be of importance to us since this will constrain the values of the mass and charge of the Gauss-Bonnet black hole. For a general discussion on higher derivative corrected black holes and their perturbative stability, see Ref \cite{moura}.

\section{QNM using WKB method}

Studying QNMs in black hole backgrounds essentially require the solution of Scr\"{o}dinger like wave equation with a particular boundary condition. We will briefly review the derivation of the wave equation in higher dimensions due to Dirac perturbations in the appendix. 
However, for most spacetime geometries, the wave equation governing the QNMs is
not exactly solvable. In our case also the wave equation so obtained is not exactly solvable and we have to look for numerical schemes for finding out the QN frequencies. Various numerical schemes have been used in the literature to find the QN frequencies, which include direct
integration of the wave equation in the frequency domain \cite{chandra}, P\"{o}schl-Teller approximation \cite{ferrari}, WKB method \cite{will,will2,iyer,kon2}, phase integral method
\cite{Andersson, and1} and continued fraction method \cite{leaver}. We will use the WKB method in our case. The WKB method has some advantages over the other semianalytic methods since it can be carried systematically to higher orders to improve the accuracy. It has been found also that sixth order WKB methods used to find black hole QNMs give almost same result as can be found out by full numerical integrations. Also, one of the advantages of using the WKB method is that it can give an analytic expression for the frequency when one uses the lowest order WKB formula in the large multipole number limit which is otherwise almost impossible to find out.

Having discussed the black hole backgrounds of our interest we now look into the tool that we will use in evaluating the QNMs, i.e. the WKB method. As we will see in the appendix, the equation we need to solve is
\bea
\left(-\frac{d^2}{dr_{\star}^2}+V_1\right)G=\omega^2G;~~~~\left(-\frac{d^2}{dr_{\star}^2}+V_2\right)F=\omega^2F,
\eea  
where $r_{\star}$ is the tortoise coordinate and the potential $V_{1,2}$ is given by 
\bea
V_{1,2}=\pm \frac{dW}{dr_{\star}}+W^2, ~~ W=\frac{\sqrt{f}}{r}\left(l+\frac{d-2}{2}\right),
\eea
with the choice of outgoing boundary conditions at the horizon and spatial infinity i.e. nothing should come in from asymptotic infinity to disturb the system and nothing should come out of the horizon.

The formula for QN frequencies using third order WKB approach is given
by \cite{will,iyer}
\be
\omega^2=[V_0+(-2V_0^{\prime\prime})^{1/2}\tilde\Lambda(n)]-i(n+\frac{1}{2})(-2V_0^{\prime\prime})^{1/2}[1+\tilde\Omega(n)].\label{freq}
\ee
where, $\tilde\Lambda=\Lambda/i$ and
$\tilde\Omega=\Omega/(n+\frac{1}{2})$ and $\Lambda$ and $\Omega$ are
given by
\bea
\Lambda(n)&=&\frac{i}{(-2V^{\prime\prime}_0)^{1/2}}\left[\frac{1}{8}\left(\frac{V^{(4)}_0}{V^{\prime\prime}_0}\right)
\left(\frac{1}{4}+\nu^2\right)-\frac{1}{288}\left(\frac{V^{(3)}_0}{V^{\prime\prime}_0}\right)^2
(7+60\nu^2)\right],\nonumber\\
\Omega(n)&=&\frac{(n+\frac{1}{2})}{(-2V^{\prime\prime}_0)^{1/2}}\bigg [\frac{5}{6912}
\left(\frac{V^{(3)}_0}{V^{\prime\prime}_0}\right)^4
(77+188\nu^2)\nonumber\\&&-
\frac{1}{384}\left(\frac{V^{(3)^2}_0V^{(4)}_0}{V^{\prime\prime^3}_0}\right)
(51+100\nu^2)
+\frac{1}{2304}\left(\frac{V^{(4)}_0}{V^{\prime\prime}_0}\right)^2(67+68\nu^2)
\nonumber\\&&+\frac{1}{288}
\left(\frac{V^{(3)}_0V^{(5)}_0}{V^{\prime\prime^2}_0}\right)(19+28\nu^2)-\frac{1}{288}
\left(\frac{V^{(6)}_0}{V^{\prime\prime}_0}\right)(5+4\nu^2)\bigg ].\label{wl}
\eea
Where, $V^{(n)}_0=(d^nV/dx^n)_{x=x_0}$ and $\nu=n+1/2$.

It may be mentioned here that the accuracy of the WKB method depends
on the multipole number $l$ and the overtone number $n$. It has been
shown \cite{cardosoyousi} that the WKB approach is a good one for
$l>n$, i.e. the numerical and the WKB results are in good agreement if
$l>n$, but the WKB approach is not so good if $l=n$ and not at all
applicable for $l<n$.

Now, in our case the form of the potential for Reissner-Nordstr\"{o}m type solutions is given by

\bea
V_1(r)&=&f\frac{d}{dr}\left\{\sqrt{f}\frac{\kappa}{r}\right\}+f\frac{\kappa^2}{r^2}, \label{potl22}\\
&=& \kappa \sqrt{\frac{\Delta}{r^{2d-6}}}\frac{1}{r^{2(d+1)}}\left[r^d\left\{(d-1)\mu r^3+r^d\left(\sqrt{\frac{\Delta}{r^{2d-6}}}\kappa-1\right)\right\}-(d-2)\theta^2r^6\right],\nonumber
\eea
where $\Delta=(r^{2d-6}-2\mu r^{d-3}+\theta^2)$. One can also write $\Delta$ as
\bea
\Delta=(r^{d-3}-r_-^{d-3})(r^{d-3}-r_+^{d-3}).
\eea
where $r_+$ and $r_-$ are the outer and inner horizons for the Reissner-Nordstr\"{o}m black hole in $d$=dimensions. The explicit form of $r_{\pm}$ is given by
\bea
r_{\pm}=(\mu\pm\sqrt{\mu^2-\theta^2})^{\frac{1}{d-3}}.
\eea
We will need this particular form for the discussion of large angular momentum limit in the next section.

The potential for the Reissner-Nordstr\"{o}m type black holes and the charged Gauss-Bonnet black hole for different dimensions are plotted in figure (1) for a particular value of multipole number, charge and Gauss-Bonnet coupling. Since the QN frequencies depend on the height and width of the potential, it is clear from the picture, at least qualitatively, that the quasinormal frequencies should be different for the two different space times discussed in the paper. Even with a very small value of the Gauss Bonnet coupling ($\alpha=0.1$ in the figure (1)), there is significant changes in the potential. 

 \begin{figure}[here]
%\begin{center}
%\centerline{\hspace{0.1mm}
\centering
\subfigure[Potential for Reissner-Nordstr\"{o}m black hole]{\rotatebox{0}{\epsfxsize=8cm\epsfbox{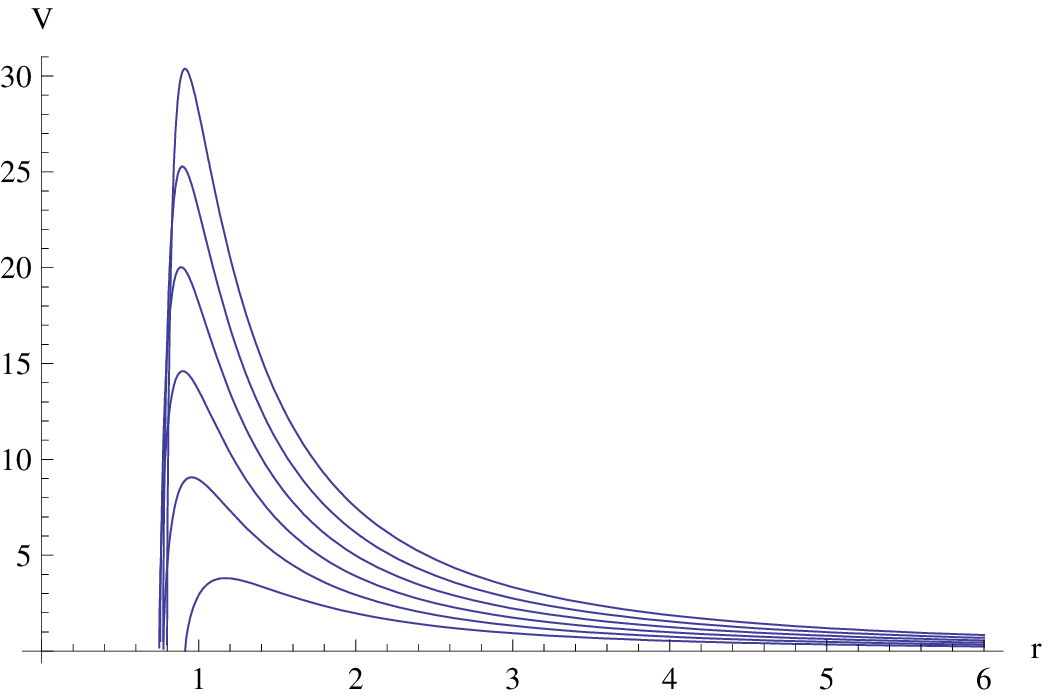}}}
%\caption[]{The potential for higher dimensional Reissner-Nordstr\"{o}m like black holes. The lower one is for $d=5$ and the topmost one is for $d=10$ for charge $Q=0.05$ and $l=2$}
\subfigure[Potential for charged Gauss-Bonnet black hole with $\alpha=0.1$]{\rotatebox{0}{\epsfxsize=8cm\epsfbox{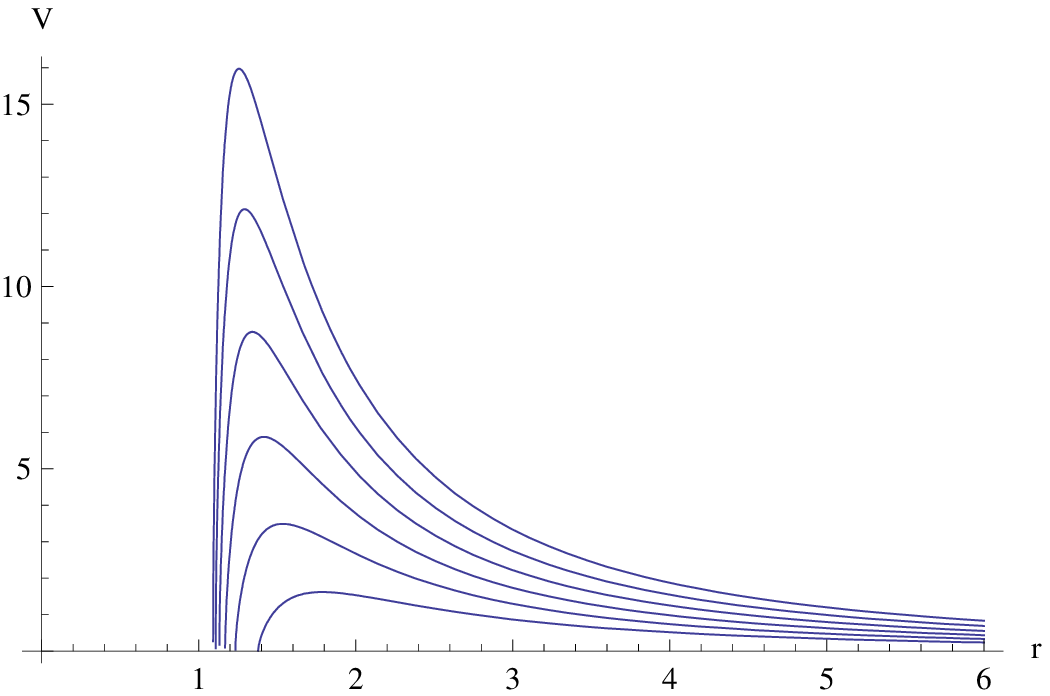}}}
%\end{center}
\caption{The potential for higher dimensional (a) Reissner-Nordstr\"{o}m and (b) charged Gauss-Bonnet black holes. The lower plot is for $d=5$ and the topmost one is for $d=10$ for charge $Q=0.05$ and $l=2$ in both cases}
\end{figure}

By properly choosing the $\Delta$ one can get the potential for the Reissner-Nordstr\"{o}m case for $d=4$, which is given in \cite{wu}. Similarly a potential for the charged Gauss Bonnet black hole can be found out. Due to the complicated expression for the potential for charged Gauss-Bonnet case, we do not explicitly write down the form here. 

Having found out the explicit form for the potential, our next goal is to determine the quasinormal frequencies for Reissner-Nordstr\"{o}m type black holes and charged Gauss-Bonnet black holes in higher dimensions. For that we will use the WKB formula for the QN frequencies given by Eqn. (\ref{freq}). In this paper we will be looking at the above mentioned black hole backgrounds in space-time dimensions $d=5$ to $d=10$. 

\begin{figure}[htp]
     \centering
     \subfigure[$d=5,~ Re(\omega)~ vs~ Q$]{\rotatebox{0}{\epsfxsize=5.1cm\epsfbox{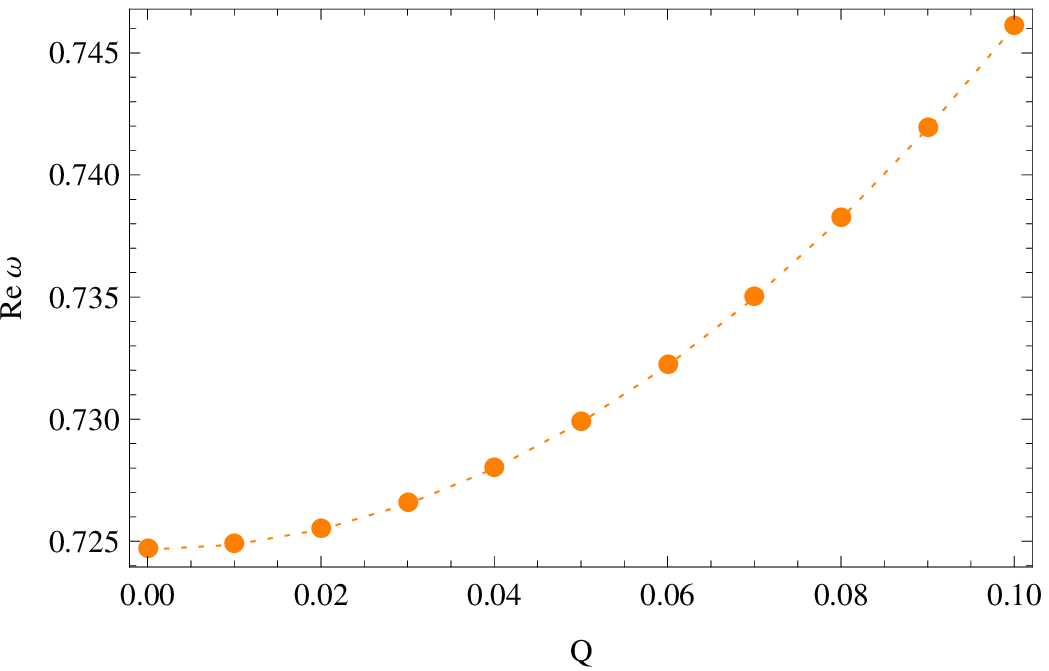}}}
     \hspace{.2in}
     \subfigure[$d=6,~ Re(\omega)~ vs~ Q$]{\rotatebox{0}{\epsfxsize=5.1cm\epsfbox{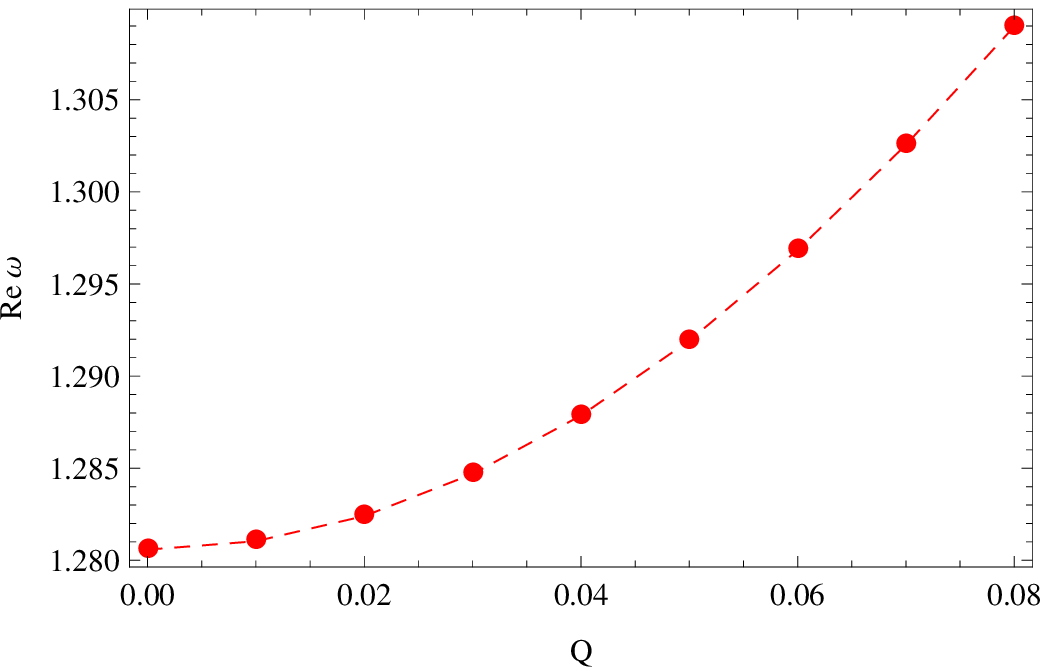}}}
%{\includegraphics[width=.15\textwidth]{d7lc.ps}}\\
     \vspace{.05in}
   \hspace{.2in}
     \subfigure[$d=7,~ Re(\omega)~ vs~ Q$]{\rotatebox{0}{\epsfxsize=5.1cm\epsfbox{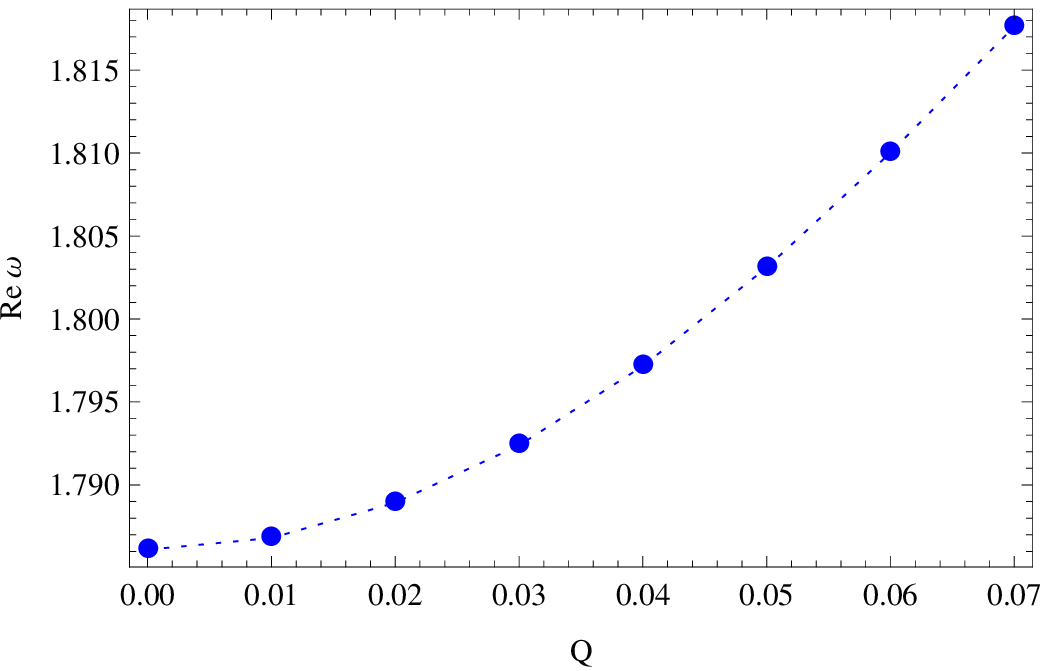}}}
%{\includegraphics[width=.15\textwidth]{d8l.ps}}
\hspace{.2in}     
\subfigure[$d=8,~ Re(\omega)~ vs~ Q$]{\rotatebox{0}{\epsfxsize=5.1cm\epsfbox{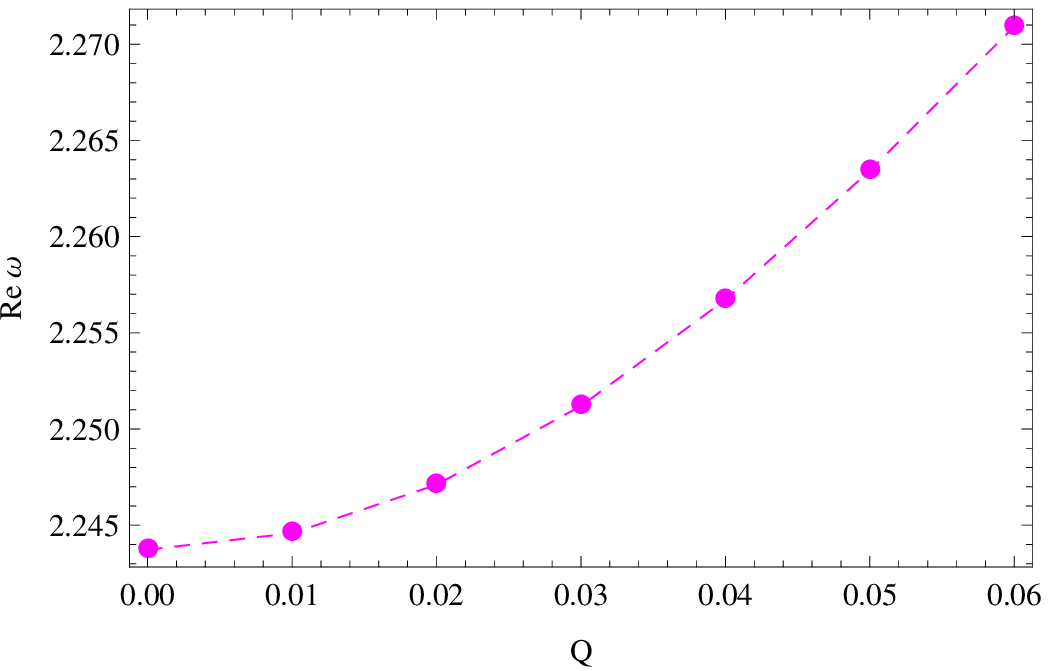}}}
     %\caption{}
\hspace{.2in}     
\subfigure[$d=9,~ Re(\omega)~ vs~ Q$]{\rotatebox{0}{\epsfxsize=5.1cm\epsfbox{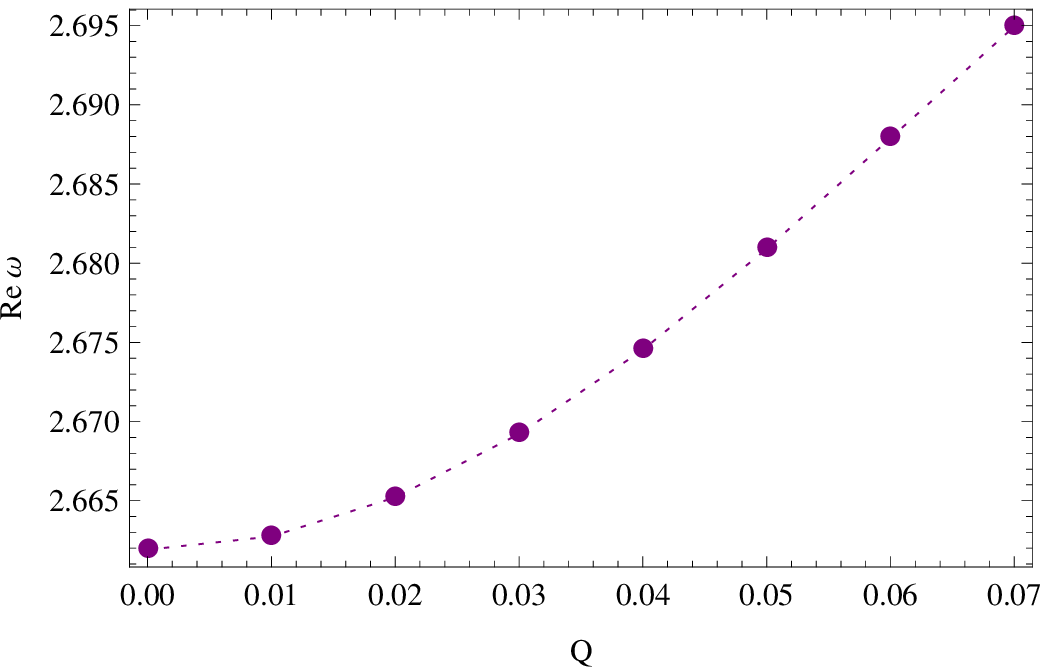}}}
     %\caption{}\hspace{.3in}     
\hspace{.2in} 
\subfigure[$d=10,~ Re(\omega)~ vs~ Q$]{\rotatebox{0}{\epsfxsize=5.1cm\epsfbox{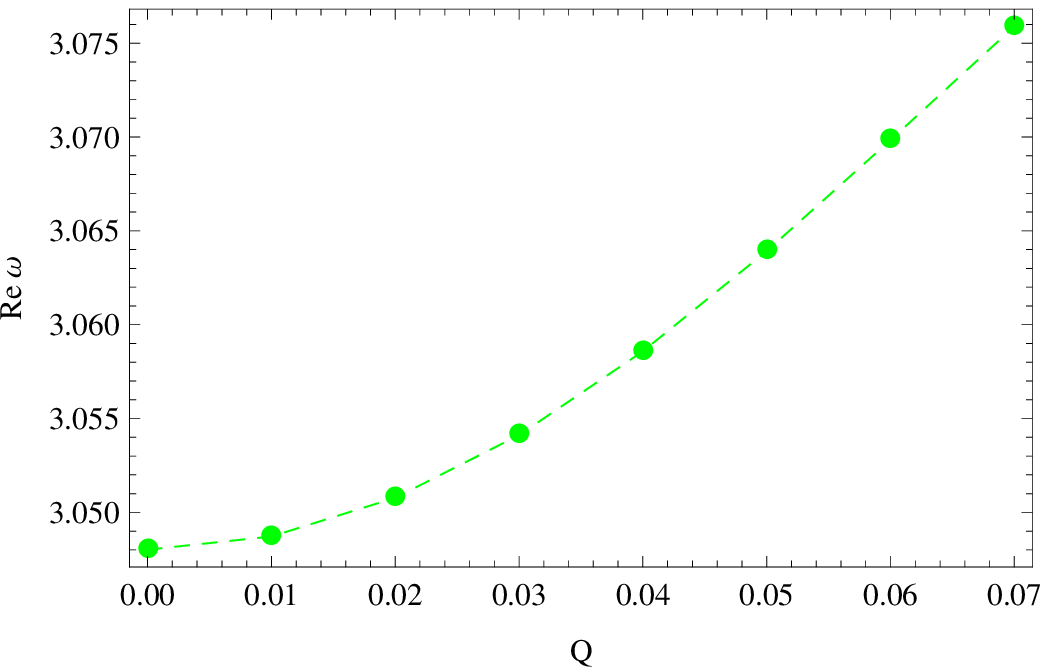}}}
\hspace{.1in}
\subfigure[$d=5,~ -Im(\omega)~ vs~ Q$]{\rotatebox{0}{\epsfxsize=5.1cm\epsfbox{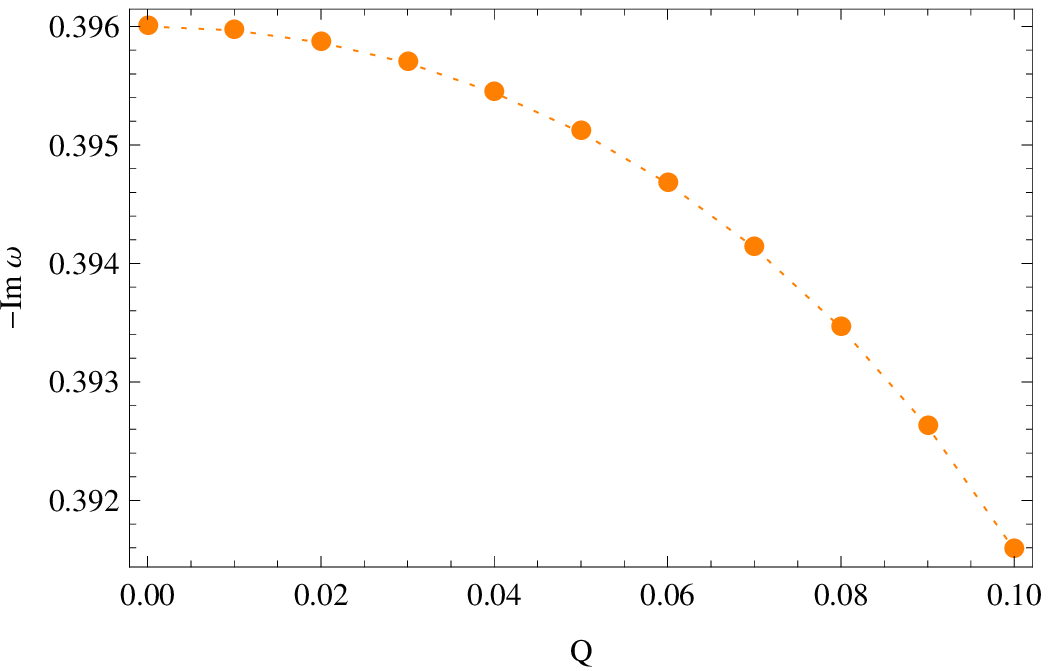}}}
     \hspace{.2in}
     \subfigure[$d=6,~ -Im(\omega)~ vs~ Q$]{\rotatebox{0}{\epsfxsize=5.1cm\epsfbox{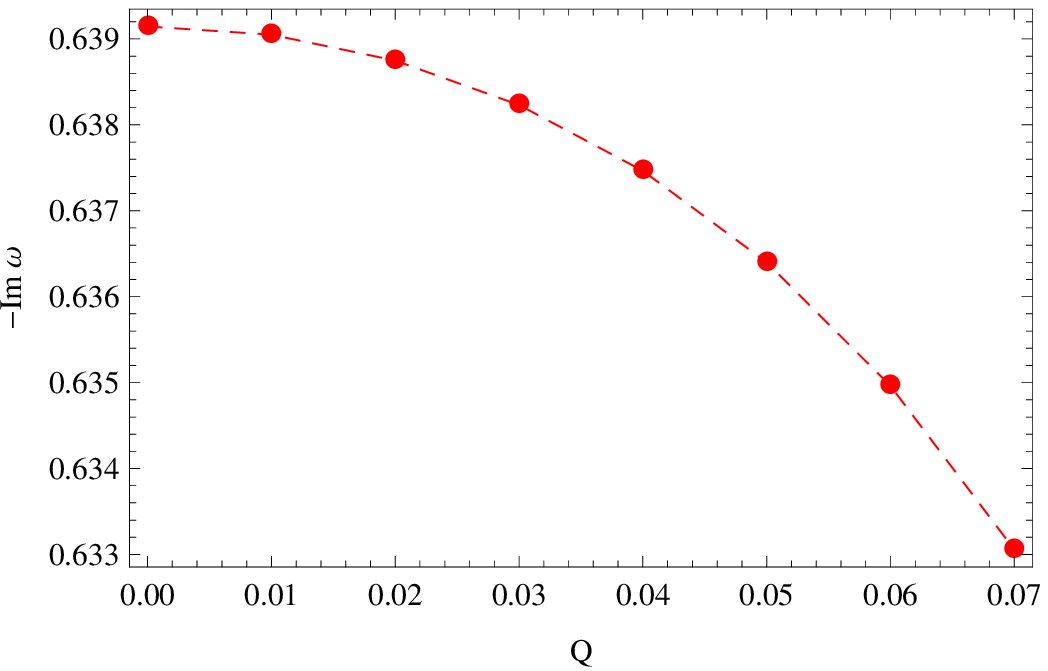}}}
%{\includegraphics[width=.15\textwidth]{d7lc.ps}}\\
     \vspace{.05in}
   \hspace{.2in}
     \subfigure[$d=7,~ -Im(\omega)~ vs~ Q$]{\rotatebox{0}{\epsfxsize=5.1cm\epsfbox{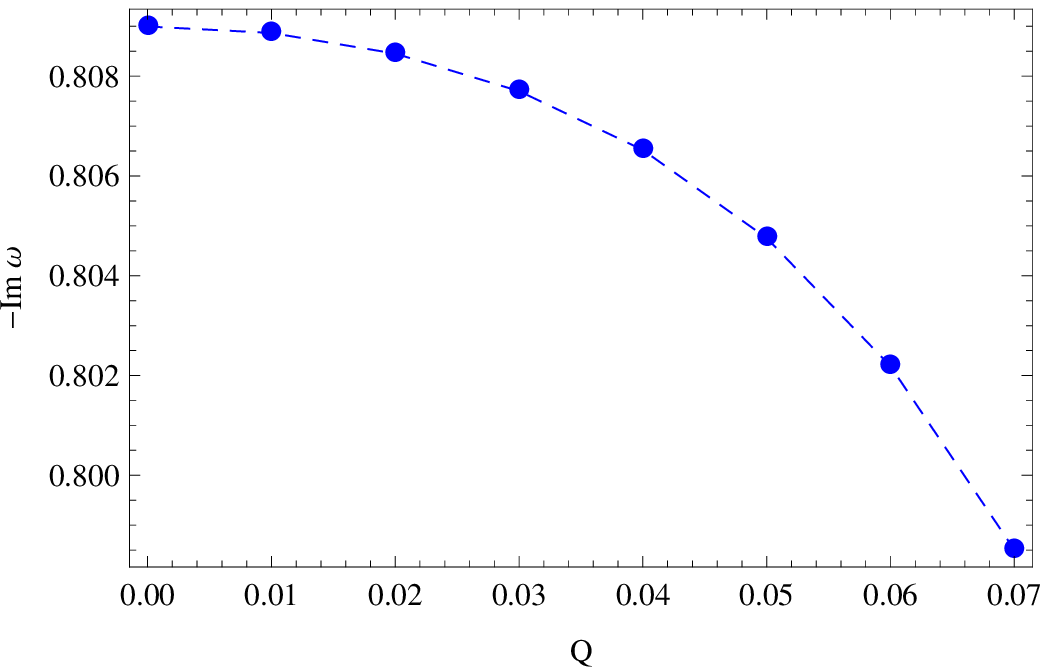}}}
%{\includegraphics[width=.15\textwidth]{d8l.ps}}
\hspace{.2in}     
\subfigure[$d=8,~ -Im(\omega)~ vs~ Q$]{\rotatebox{0}{\epsfxsize=5.1cm\epsfbox{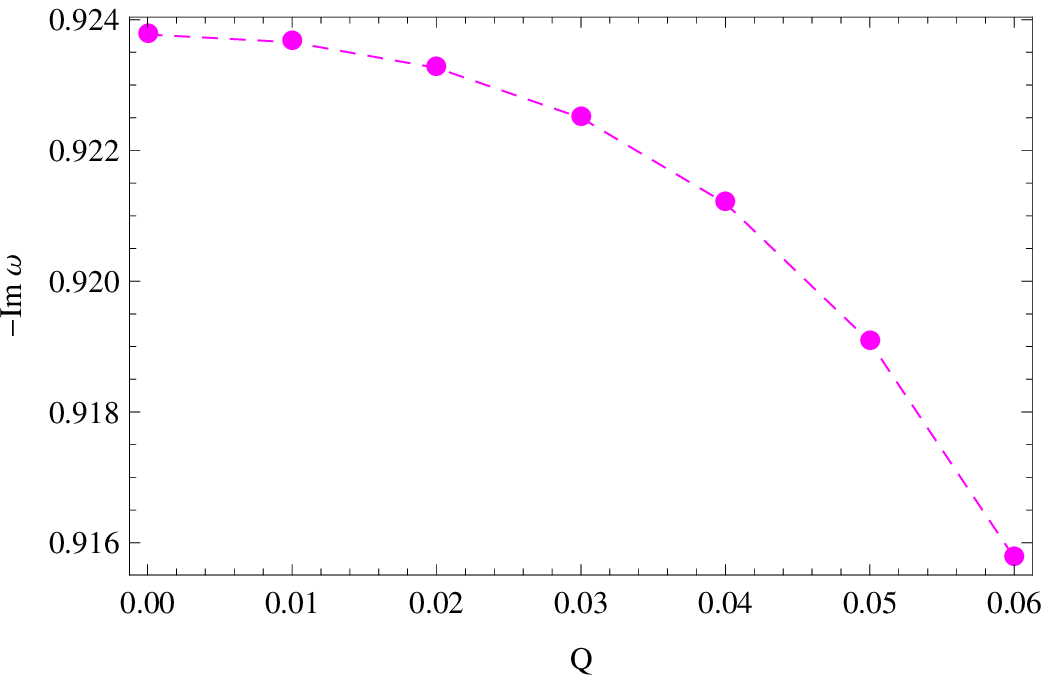}}}
     %\caption{}
\hspace{.2in}     
\subfigure[$d=9,~ -Im(\omega)~ vs~ Q$]{\rotatebox{0}{\epsfxsize=5.1cm\epsfbox{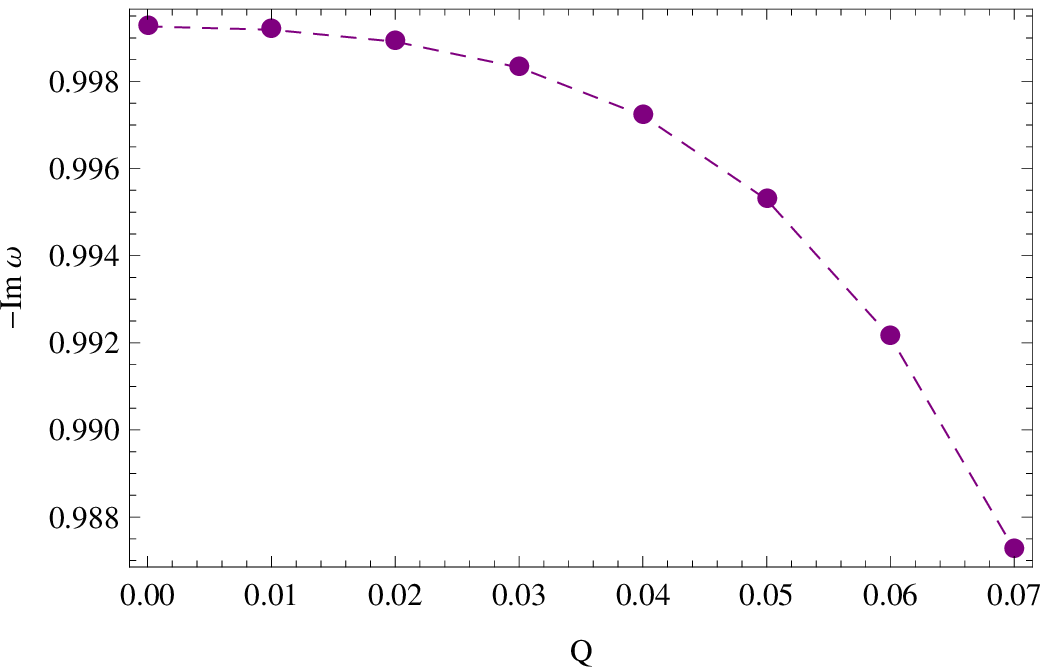}}}
     %\caption{}\hspace{.3in}     
\hspace{.2in} 
\subfigure[$d=10,~ -Im(\omega)~ vs~ Q$]{\rotatebox{0}{\epsfxsize=5.1cm\epsfbox{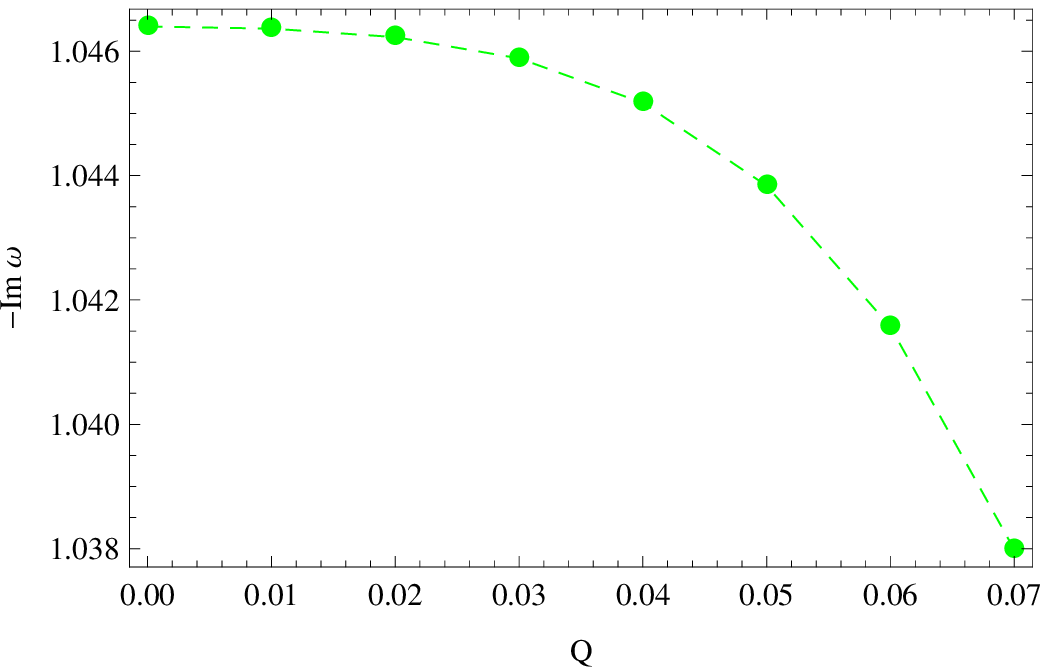}}}     
\caption{The behaviour of Real (a-f) and Imaginary (g-l) part of the quasinormal frequency $\omega$ with charge $Q$ for 
Reissner-Nordstr\"{o}m type black hole in $d=5$ to $10$ for $l=0$ and $n=0$.}
\label{fig2}
\end{figure}

In figure (\ref{fig2}) we show the behaviour of the real (2(a)-2(f)) and imaginary part (2(g)-2(l)) of the QN frequency $\omega$ with charge $Q$ for Reissner-Nordstr\"{o}m type solutions in higher dimensions. The plots are made for a particular value of the multipole number $l=0$ and mode number $n=0$. As can be seen from the figure the real part of the frequency increases with the increase of charge. This implies the real oscillation frequency increases as the charge is increased. 

\begin{figure}[htp]
\centering
\subfigure[Variation of Re $\omega$ with $Q$ and $l$]{\rotatebox{0}{\epsfxsize=8.2cm\epsfbox{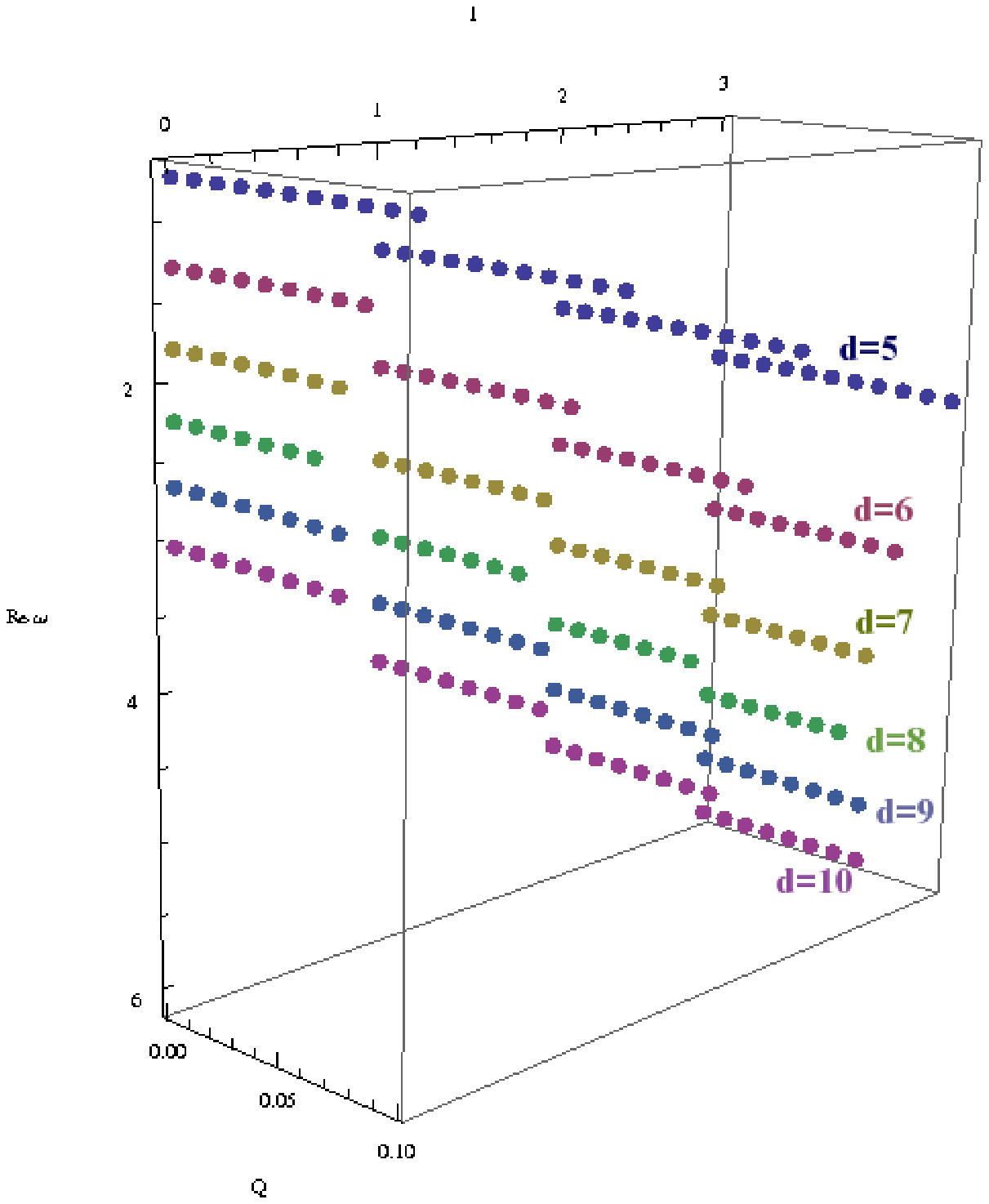}}}
\hspace{0.2cm}%\caption[]{$\mathbb{I}$m $\omega$ as a function of Gauss-Bonnet
\subfigure[Variation of Im $\omega$ with $Q$ and $l$]{\rotatebox{0}{\epsfxsize=7.6cm\epsfbox{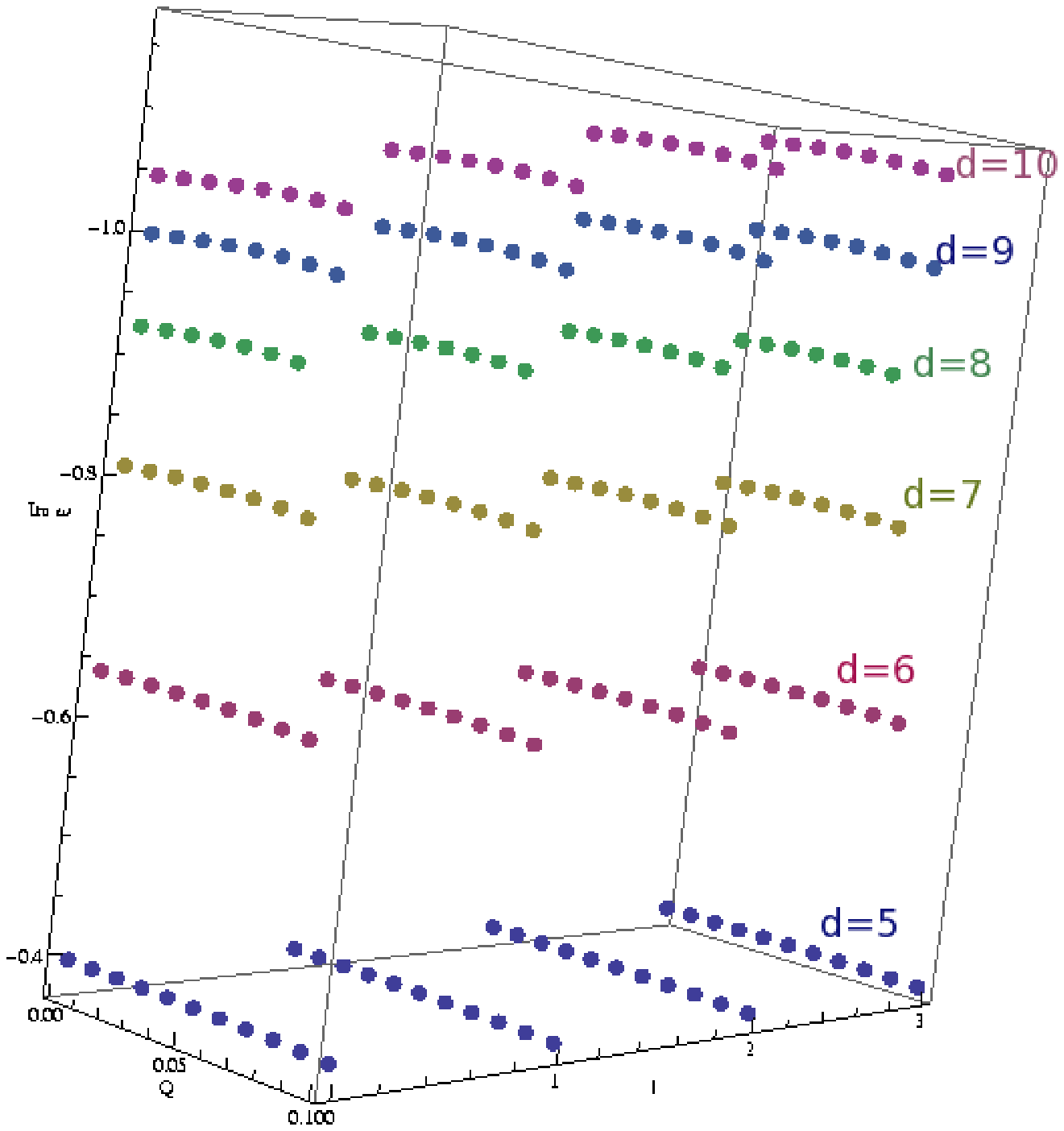}}}
\caption{Behaviour of the (a) Real and (b) Imaginary part of $\omega$ with $l$ and $Q$ for the higher dimensional RN black hole}
\label{fig4}
\end{figure}

% \begin{figure}[htp]
%     \centering
%     \subfigure[$d=5,~ -Im(\omega)~ vs~ Q$]{\rotatebox{0}{\epsfxsize=5.1cm\epsfbox{chgd5l0Im.eps}}}
%     \hspace{.2in}
%     \subfigure[$d=6,~ -Im(\omega)~ vs~ Q$]{\rotatebox{0}{\epsfxsize=5.1cm\epsfbox{chgd6l0Im.eps}}}
%%{\includegraphics[width=.15\textwidth]{d7lc.ps}}\\
%     \vspace{.05in}
%   \hspace{.2in}
%     \subfigure[$d=7,~ -Im(\omega)~ vs~ Q$]{\rotatebox{0}{\epsfxsize=5.1cm\epsfbox{chgd7l0Im.eps}}}
%%{\includegraphics[width=.15\textwidth]{d8l.ps}}
%\hspace{.2in}     
%\subfigure[$d=8,~ -Im(\omega)~ vs~ Q$]{\rotatebox{0}{\epsfxsize=5.1cm\epsfbox{chgd8l0Im.eps}}}
%     %\caption{}
%\hspace{.2in}     
%\subfigure[$d=9,~ -Im(\omega)~ vs~ Q$]{\rotatebox{0}{\epsfxsize=5.1cm\epsfbox{chgd9l0Im.eps}}}
%     %\caption{}\hspace{.3in}     
%\hspace{.2in} 
%\subfigure[$d=10,~ -Im(\omega)~ vs~ Q$]{\rotatebox{0}{\epsfxsize=5.1cm\epsfbox{chgd10l0Im.eps}}}
%     \caption{The behaviour of negative Imaginary part of the quasinormal frequency $\omega$ with charge 
%$Q$ for Reissner-Nordstr\"{o}m like black hole in $d=5$ to $10$ for $l=0$ and $n=0$.}
%\label{fig3}
%\end{figure}
%
The behaviour of the imaginary part of the QN frequency $\omega$ with charge $Q$ for Reissner-Nordstr\"{o}m type solutions in higher dimensions for $l=0$ and $n=0$ is shown in Fig. (2(g)-2(l)). As can be seen from the figure the negative imaginary part of the frequency falls off with the increase of charge. This implies that the damping decreases with the increase of the charge, i.e. a longer ringdown phase. For a better and compact look at the behaviour of the frequencies with charge and multipole number see Fig. (\ref{fig4}) 

It may be mentioned here that, in the mathematica code, if we use $d=4$, then the values for the QN frequencies in four space time dimensions can be found out and our result matches with the result obtained by \cite{wu} and \cite{jing3}. Both the authors of \cite{wu, jing3} have used P\"{o}schl Teller approximation scheme to determine the QN frequencies whereas we have used WKB method to determine the same. We have also checked our results with the results obtained by Cho et al in \cite{split} by explicitly putting $Q=0$ in the metric and have found that the results matches with the results of higher dimensional Schwarzschild black holes as obtained by them.   

\begin{figure}[htp]
\centering
\subfigure[Variation of Re $\omega$ with $d$]{\rotatebox{0}{\epsfxsize=7.8cm\epsfbox{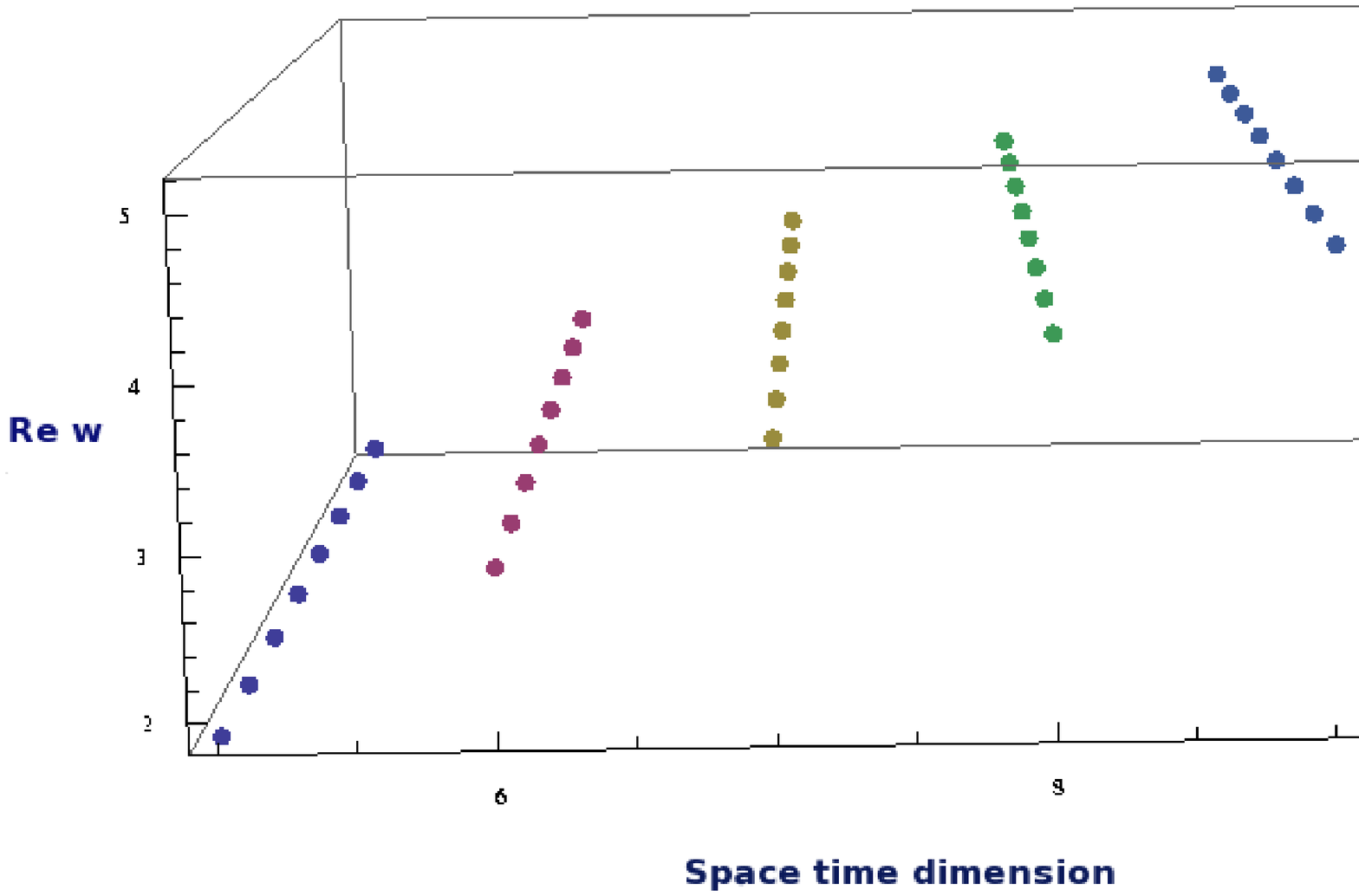}}}
\hspace{0.2cm}%\caption[]{$\mathbb{I}$m $\omega$ as a function of Gauss-Bonnet
\subfigure[Variation of Im $\omega$ with $d$]{\rotatebox{0}{\epsfxsize=7.8cm\epsfbox{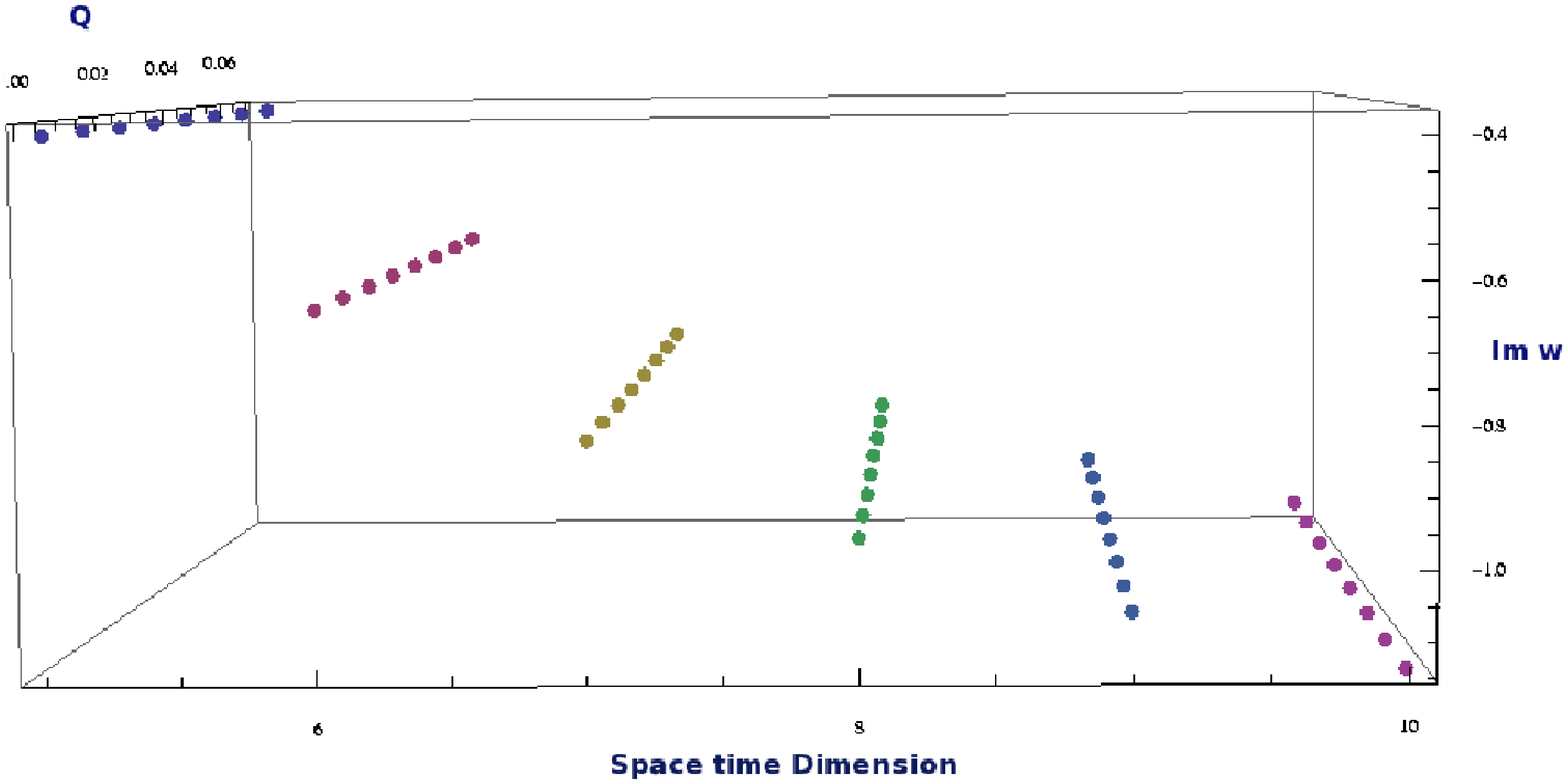}}}
\caption{Behaviour of the (a) Real and (b) Imaginary part of $\omega$ with space time dimension for $l=2$ and $n=0$ for the higher dimensional RN black hole}
\label{fig5a}
\vspace{0.5cm}     
\subfigure[$d=5,~ Re(\omega)~ vs~ Im(\omega)$]{\rotatebox{0}{\epsfxsize=5.1cm\epsfbox{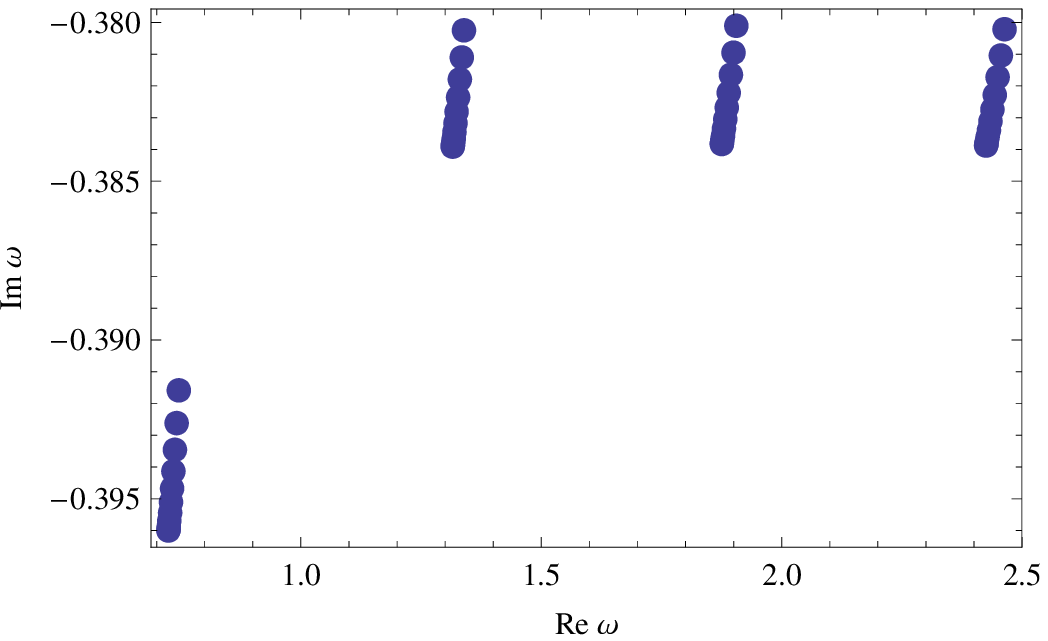}}}
     \hspace{.2in}
     \subfigure[$d=6,~ Re(\omega)~ vs~ Im(\omega)$]{\rotatebox{0}{\epsfxsize=5.1cm\epsfbox{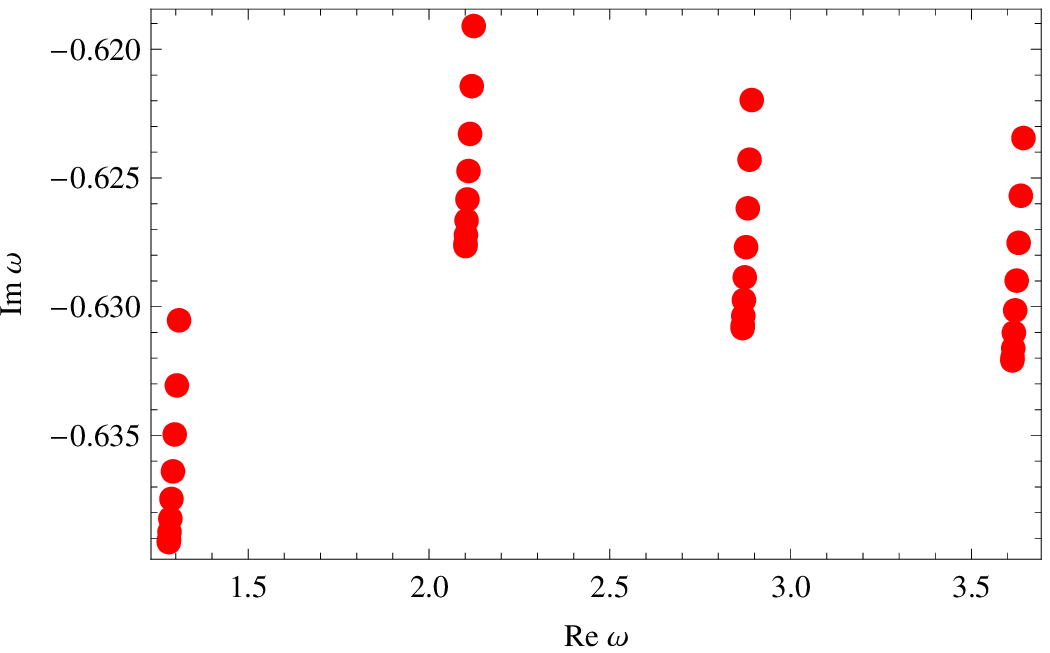}}}
%{\includegraphics[width=.15\textwidth]{d7lc.ps}}\\
     \vspace{.05in}
   \hspace{.2in}
     \subfigure[$d=7,~ Re(\omega)~ vs~ Im(\omega)$]{\rotatebox{0}{\epsfxsize=5.1cm\epsfbox{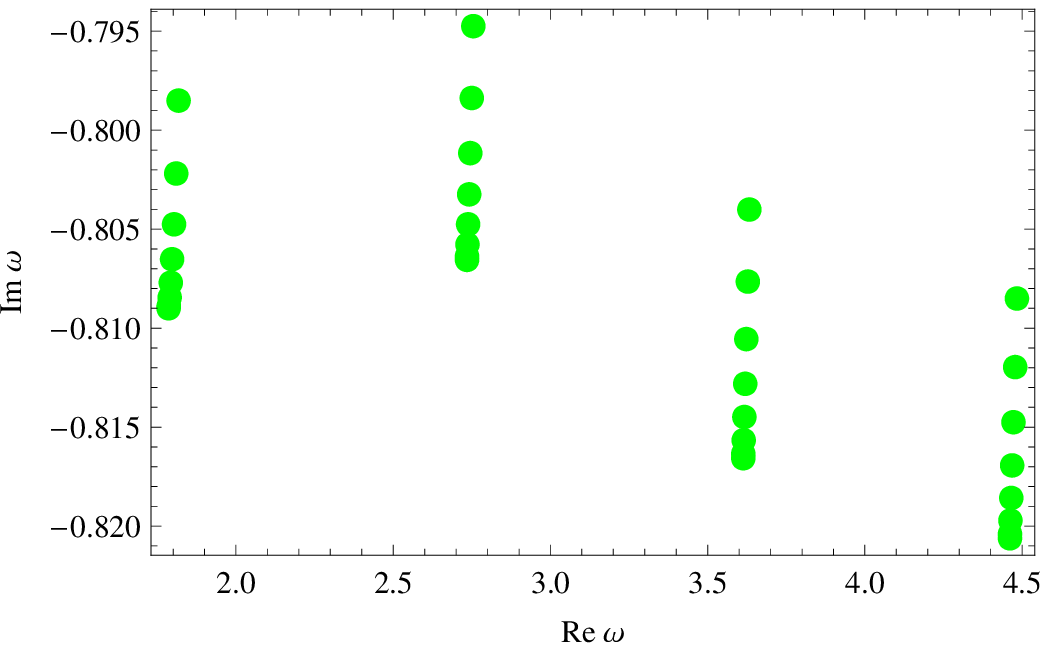}}}
%{\includegraphics[width=.15\textwidth]{d8l.ps}}
\hspace{.2in}     
\subfigure[$d=8,~ Re(\omega)~ vs~ Im(\omega)$]{\rotatebox{0}{\epsfxsize=5.1cm\epsfbox{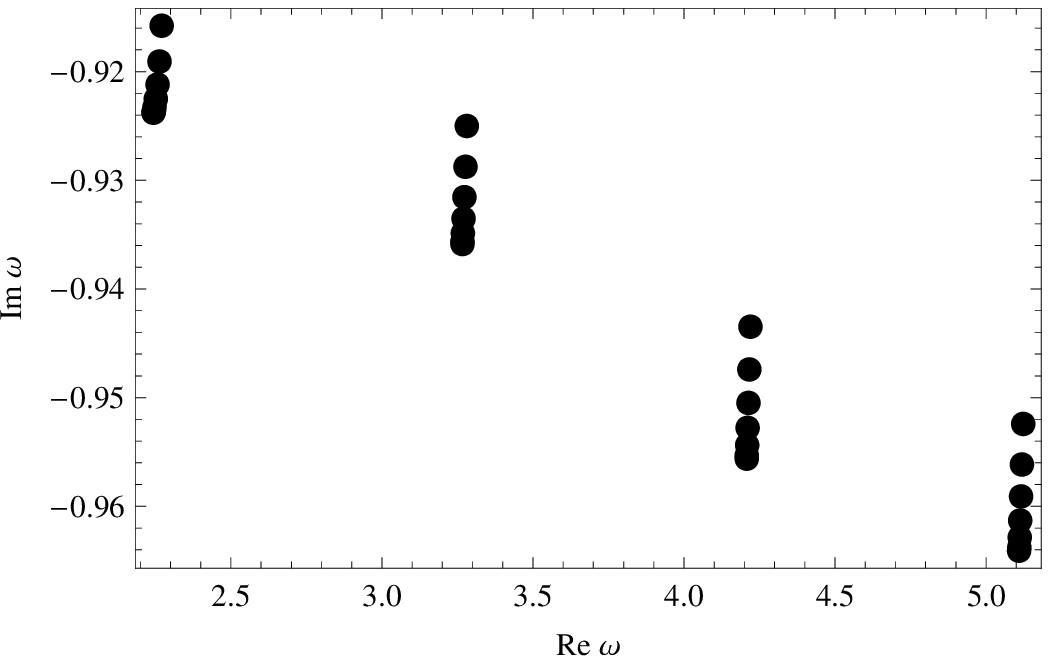}}}
     %\caption{}
\hspace{.2in}     
\subfigure[$d=9,~ Re(\omega)~ vs~ Im(\omega)$]{\rotatebox{0}{\epsfxsize=5.1cm\epsfbox{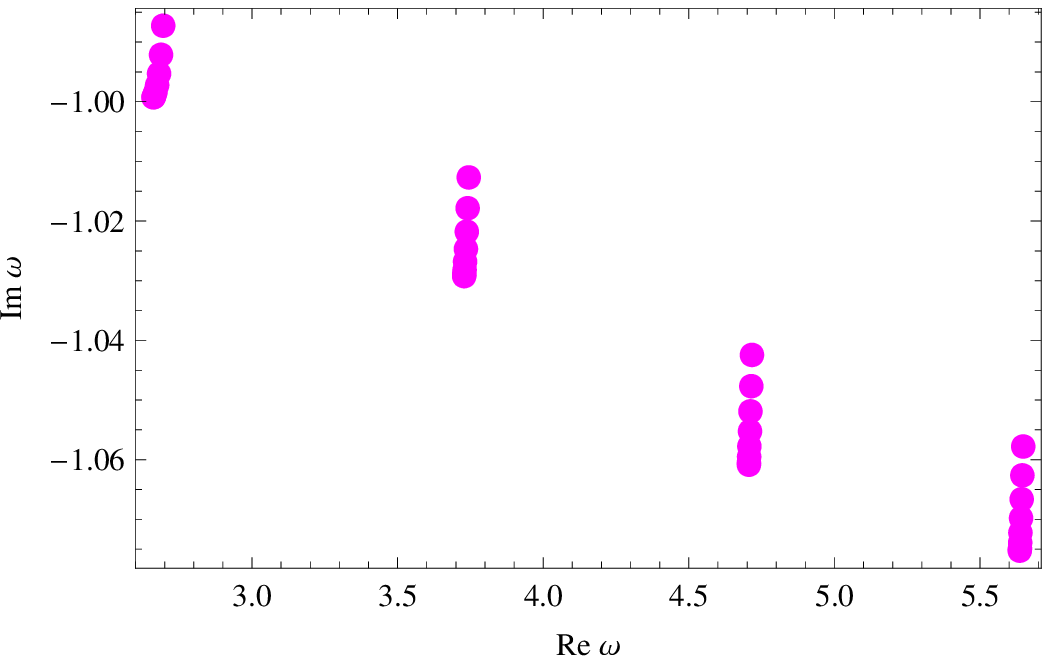}}}
     %\caption{}\hspace{.3in}     
\hspace{.2in} 
\subfigure[$d=10,~ Re(\omega)~ vs~ Im(\omega)$]{\rotatebox{0}{\epsfxsize=5.1cm\epsfbox{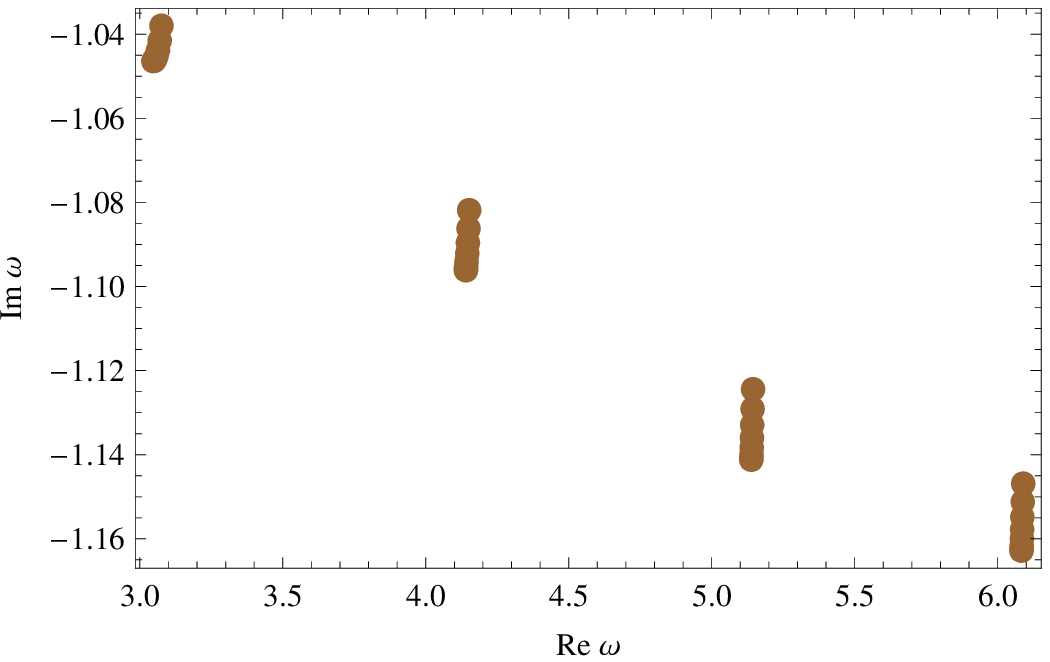}}}
     \caption{Re $\omega$ vs Im $\omega$ for different $l$ for the higher dimensional RN black hole. Each four segments in the plots correspond to 
$l=0,1,2,3$ respectively.}
\label{fig5}
\end{figure}

Let us now discuss the behaviour of the real and imaginary part of the QN frequency with the space time dimensions. The real part of the frequency increases with the increase of dimension while the imaginary part also increases with the space time dimension (See Fig. \ref{fig5a}). So, the real oscillation frequencies are increasing as the space time dimensions increase, but the damping also becomes larger for higher dimensional charged black holes. Thus it is found that the qualitative behaviour of the QN frequencies for purely higher dimensional RN black hole is the same with the behaviour of QN frequencies for higher dimensional Reissner-Nordstr\"{o}m black holes projected on the brane \cite{kanti1}. Though it may be mentioned that the quantitative nature of the frequencies are hugely different. Both the real and imaginary parts of the pure $d$-dimensional RN black holes are larger in magnitude when compared with the results of \cite{kanti1}, where the QN frequencies for RN black holes projected on the brane were calculated. We have checked this for dimensions $d=5, 6$ (for which the data were available in \cite{kanti1}).   

The real and imaginary part of the quasinormal frequency is plotted in figure (\ref{fig5}). Each four distinct segments in all the plots correspond to multipole values $l=0$, $1$, $2$ and $3$ respectively. The points in each segment correspond to different values of charge starting from $Q=0$ to $Q_{ex}$, where $Q_{ex}$ is the extremal value of the charge. 

Now, let us take a look at the results for the charged Gauss-Bonnet black hole. Due to a huge set of data, we do not give the data for all the dimensions with all values of charge, coupling and multipole number in this paper, however we only consider a particular $d=7$ and a particular multipole number $l=1$ and overtone number $n=0$ for the sake of simplicity. The study of scalar field evolution in the Gauss Bonnet background was done in \cite{konoplyaGB} and \cite{konabdalla} while vector and tensor perturbation was done in \cite{sayan} (for recent study see \cite{konGBz}). It may be noted that there were no such studies on QNMs in the charged Gauss-Bonnet background due to Dirac perturbations, this paper tries to fill in this gap in the literature.

\begin{figure}[htp]
     \centering
     \subfigure[$Re(\omega)~ vs~ Q~~for ~\alpha=0.1$]{\rotatebox{0}{\epsfxsize=5.1cm\epsfbox{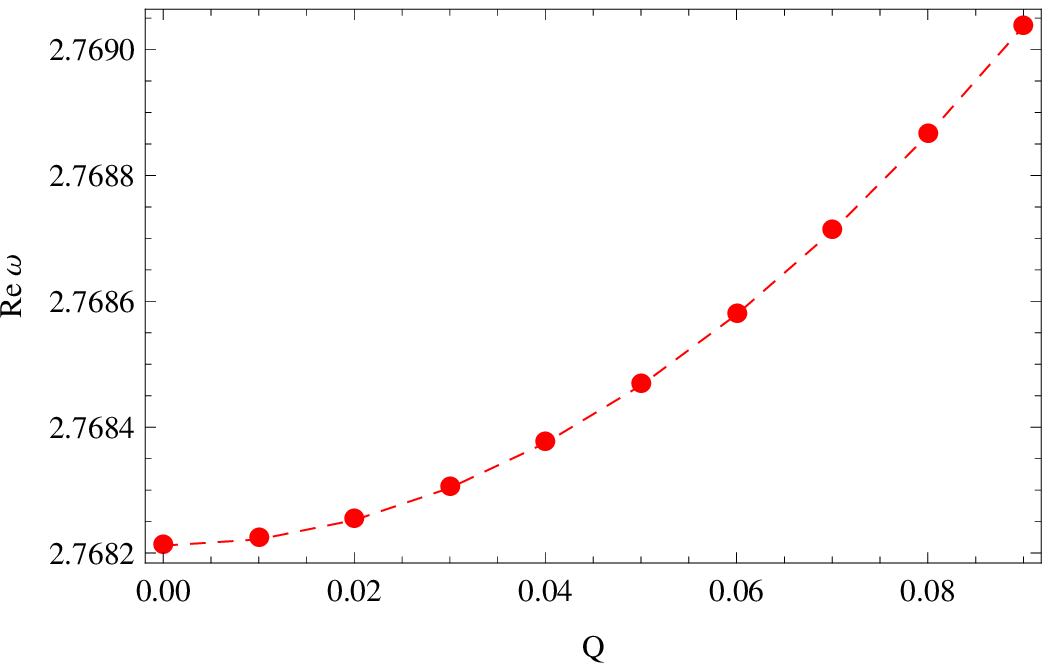}}}
     \hspace{.2in}
     \subfigure[$Re(\omega)~ vs~ Q~~for ~\alpha=0.2$]{\rotatebox{0}{\epsfxsize=5.1cm\epsfbox{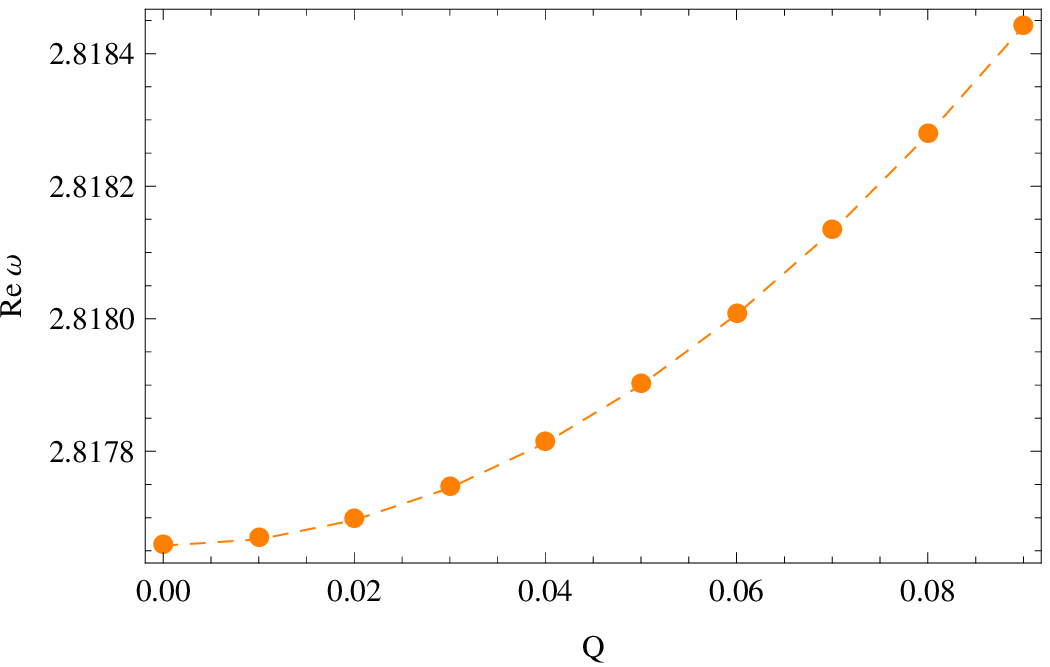}}}
%{\includegraphics[width=.15\textwidth]{d7lc.ps}}\\
     \vspace{.05in}
   \hspace{.2in}
     \subfigure[$Re(\omega)~ vs~ Q~~for ~\alpha=0.3$]{\rotatebox{0}{\epsfxsize=5.1cm\epsfbox{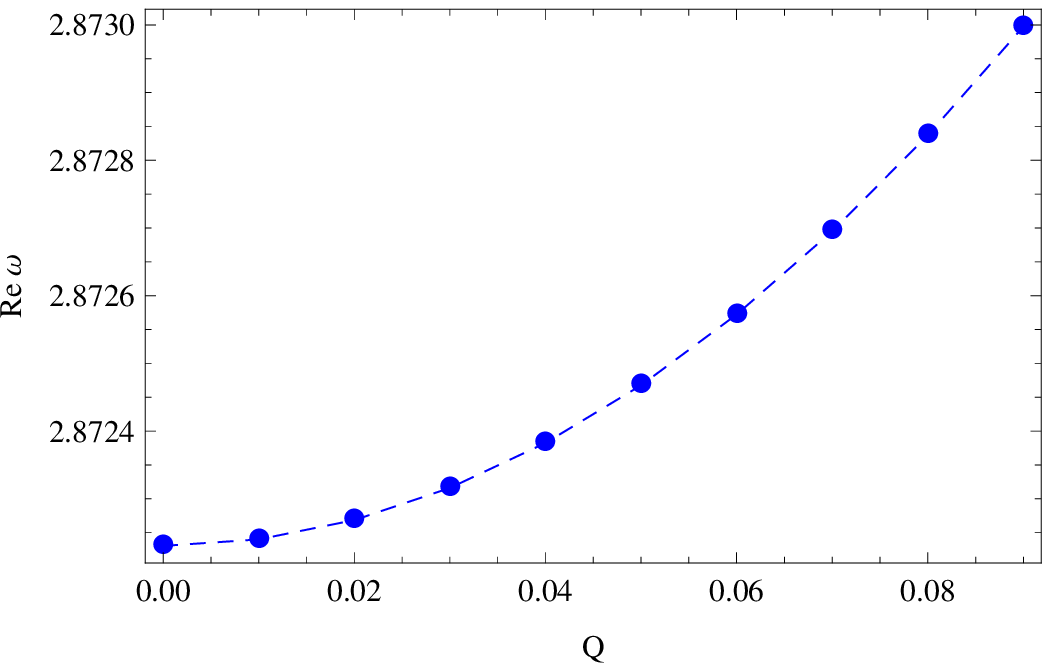}}}
%{\includegraphics[width=.15\textwidth]{d8l.ps}}
\hspace{.2in}     
\subfigure[$Re(\omega)~ vs~ Q ~for~\alpha=0.4$]{\rotatebox{0}{\epsfxsize=5.1cm\epsfbox{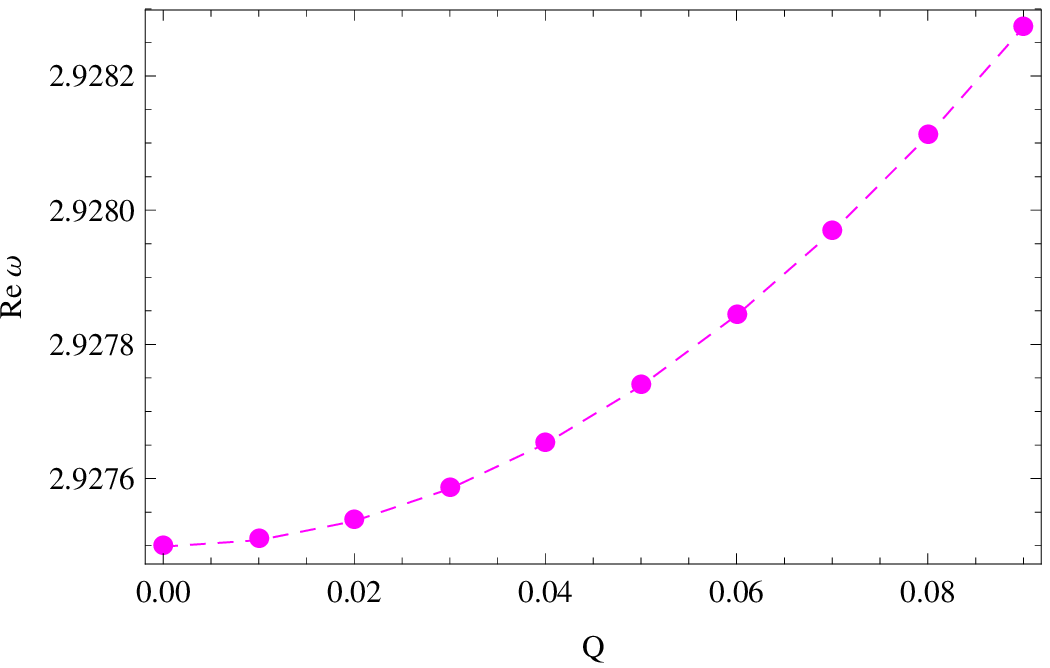}}}
     %\caption{}
\hspace{.2in}     
\subfigure[$Re(\omega)~ vs~ Q~~for ~\alpha=0.5$]{\rotatebox{0}{\epsfxsize=5.1cm\epsfbox{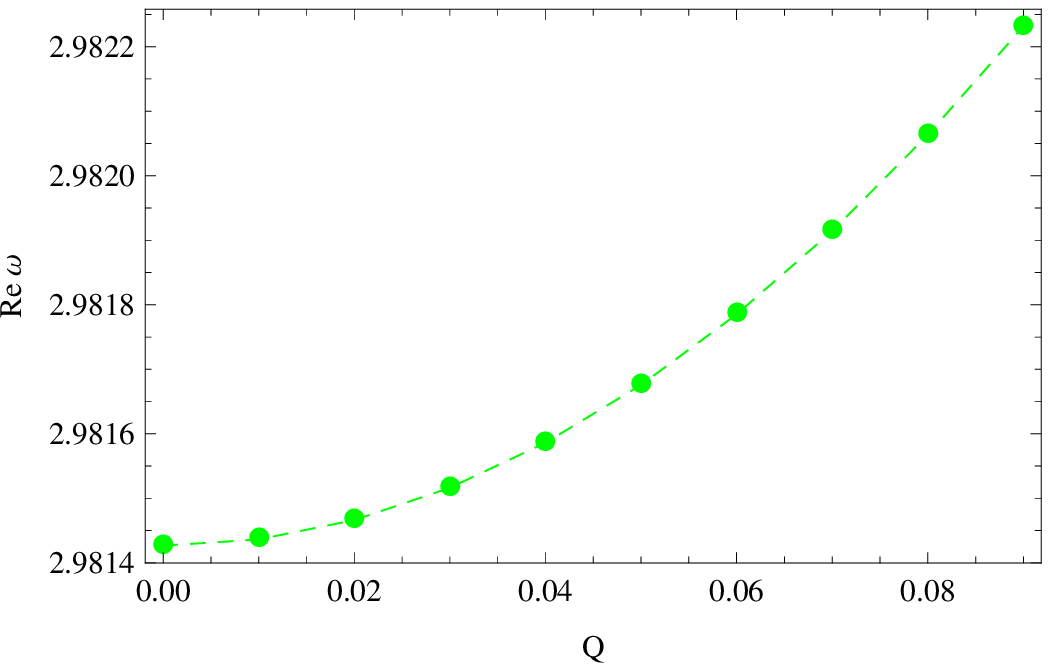}}}
     %\caption{}\hspace{.3in}     
\hspace{.2in} 
\subfigure[$Re(\omega)~ vs~ Q~for~ \alpha=1.0$]{\rotatebox{0}{\epsfxsize=5.1cm\epsfbox{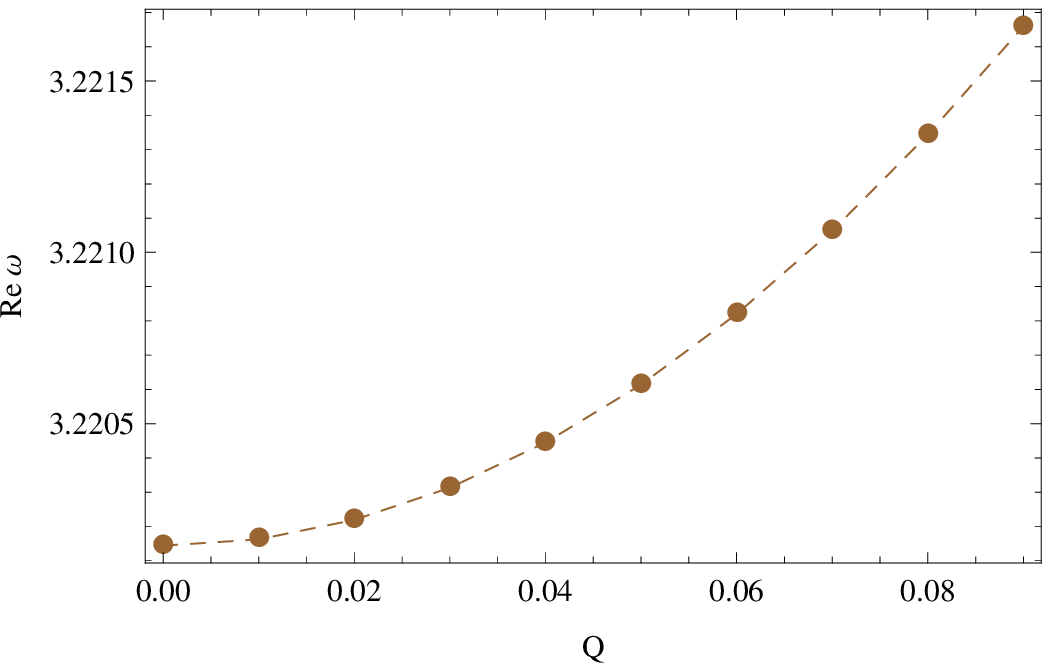}}}
\hspace{.1in}
\subfigure[$-Im(\omega)~ vs~ Q~for~ \alpha=0.1$]{\rotatebox{0}{\epsfxsize=5.1cm\epsfbox{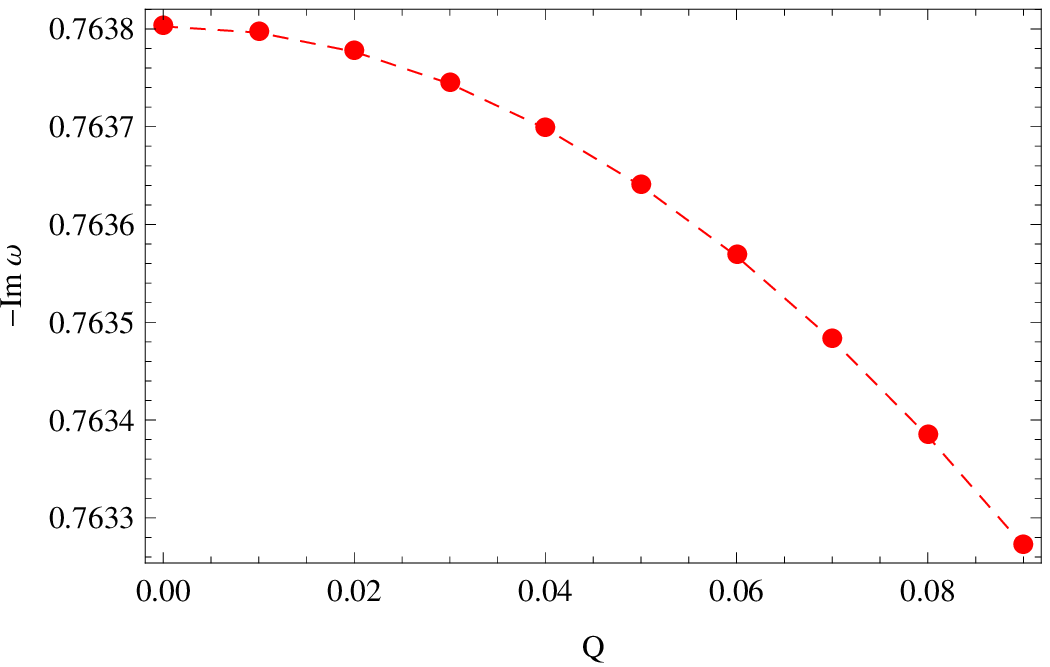}}}
     \hspace{.2in}
     \subfigure[$-Im(\omega)~ vs~ Q~for~ \alpha=0.2$]{\rotatebox{0}{\epsfxsize=5.1cm\epsfbox{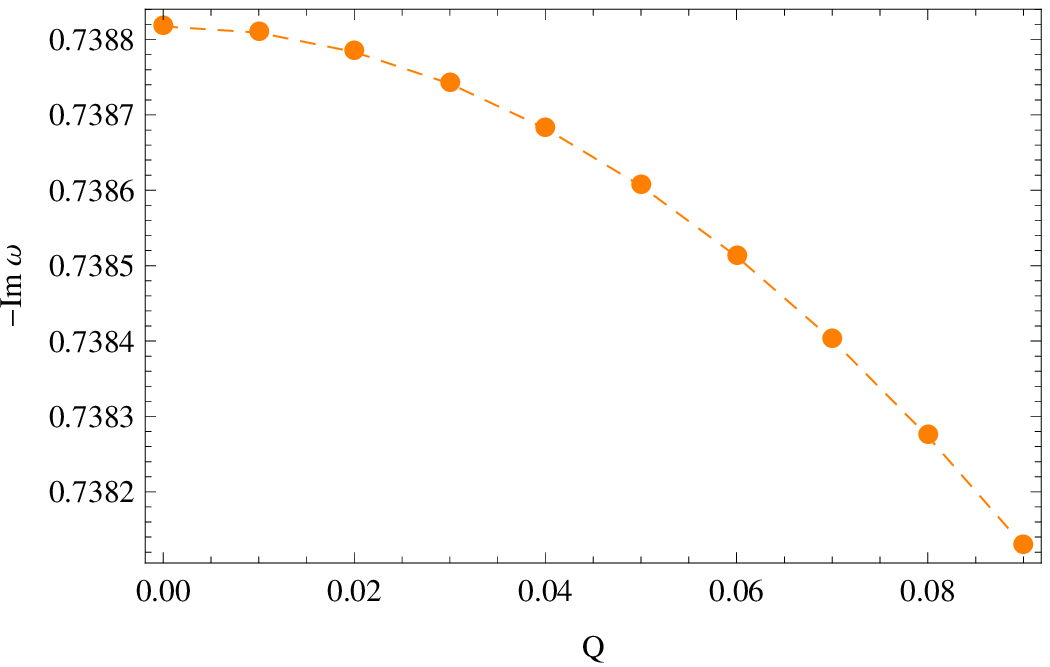}}}
%{\includegraphics[width=.15\textwidth]{d7lc.ps}}\\
     \vspace{.05in}
   \hspace{.2in}
     \subfigure[$-Im(\omega)~ vs~ Q~for~ \alpha=0.3$]{\rotatebox{0}{\epsfxsize=5.1cm\epsfbox{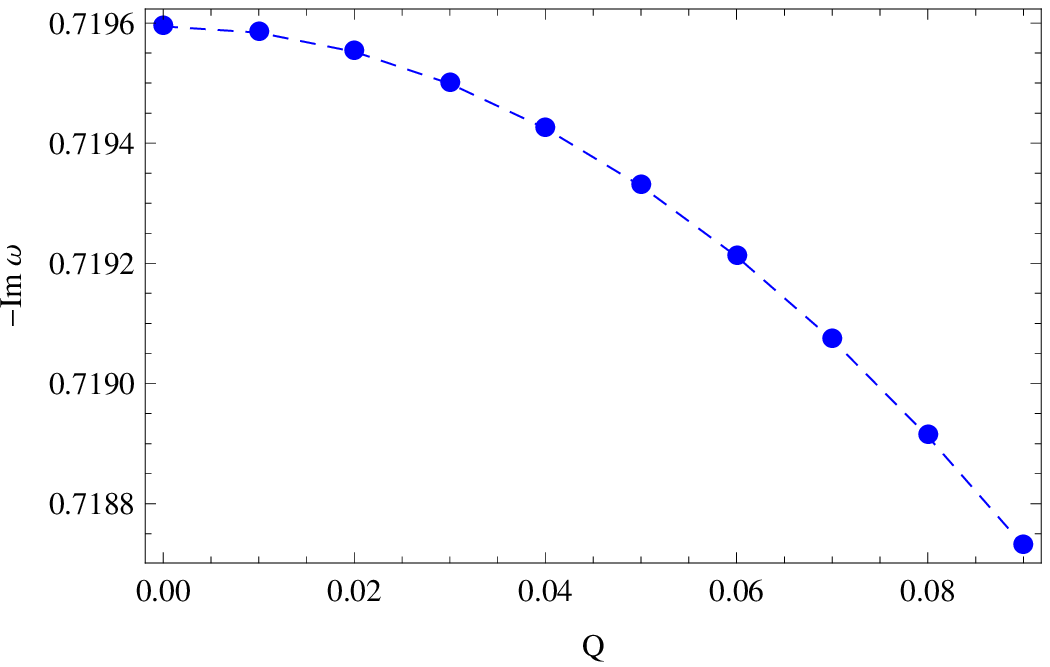}}}
%{\includegraphics[width=.15\textwidth]{d8l.ps}}
\hspace{.2in}     
\subfigure[$-Im(\omega)~ vs~ Q~for~ \alpha=0.4$]{\rotatebox{0}{\epsfxsize=5.1cm\epsfbox{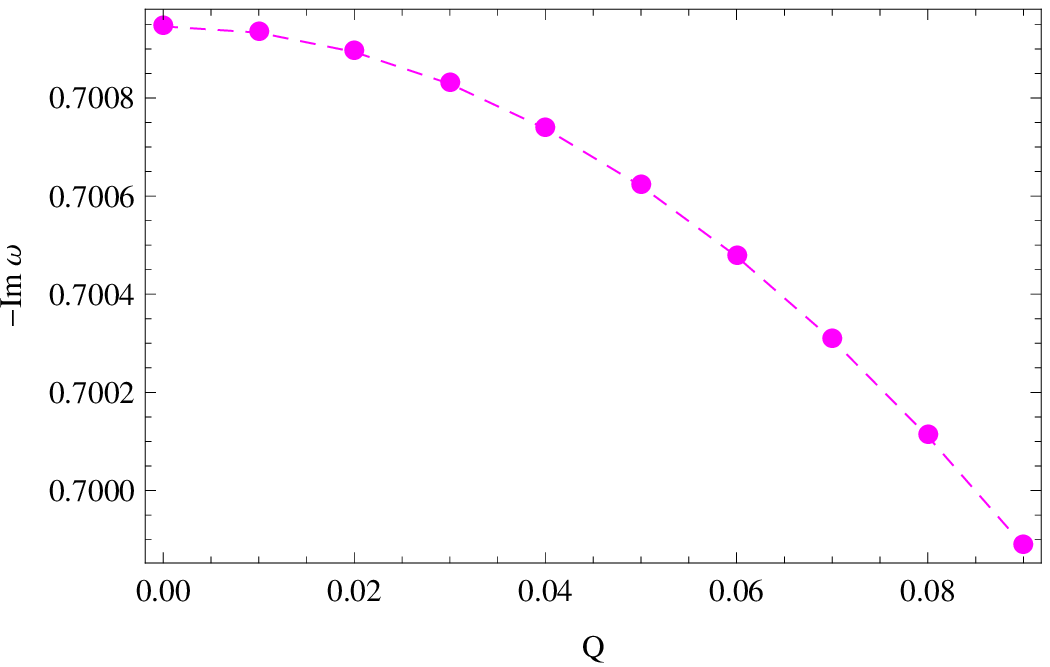}}}
     %\caption{}
\hspace{.2in}     
\subfigure[$-Im(\omega)~ vs~ Q~for~ \alpha=0.5$]{\rotatebox{0}{\epsfxsize=5.1cm\epsfbox{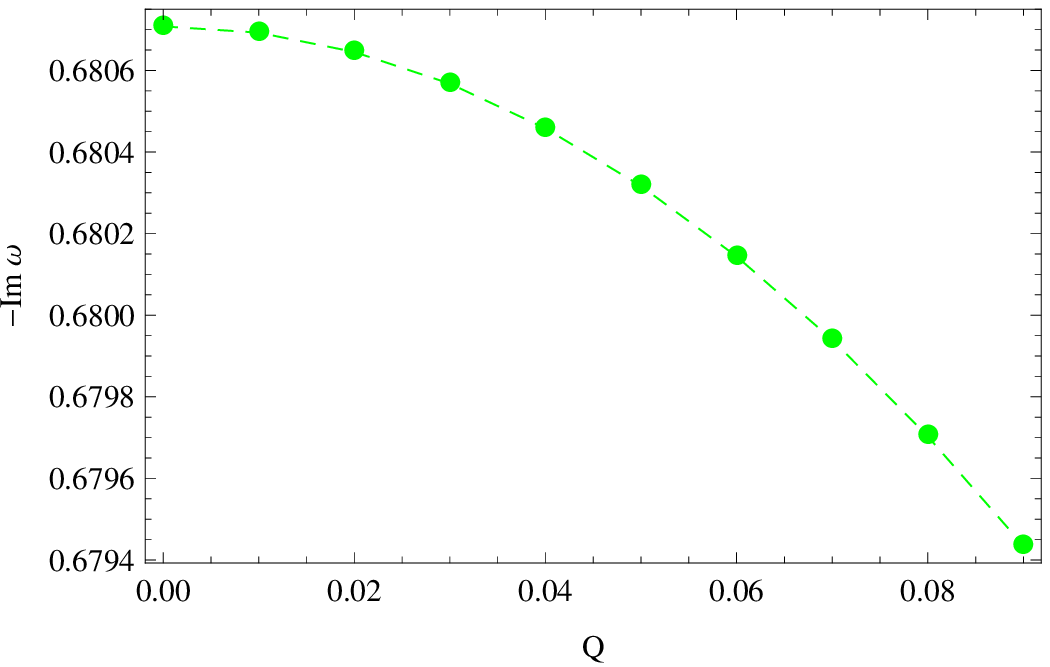}}}
     %\caption{}\hspace{.3in}     
\hspace{.2in} 
\subfigure[$-Im(\omega)~ vs~ Q~for~ \alpha=1.0$]{\rotatebox{0}{\epsfxsize=5.1cm\epsfbox{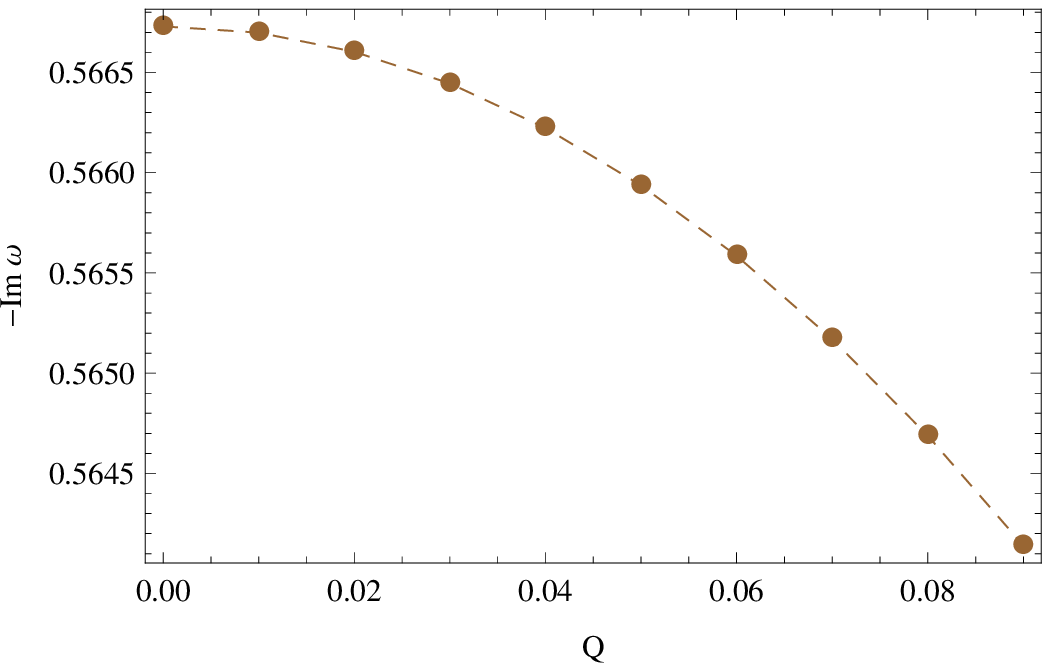}}}     
\caption{The behaviour of Real (a-f) and Imaginary (g-l) part of the quasinormal frequency $\omega$ with charge $Q$ for charged Gauss-Bonnet black hole in $d=7$ for $l=1$ and $n=0$.}
\label{fig6}
\end{figure}

As can be seen from the Figs. (\ref{fig6}(a)-\ref{fig6}(f)), the real part of the frequency for the charged Gauss-Bonnet black hole increases with the increase of charge whereas Figs. (\ref{fig6}(g)-\ref{fig6}(l)) suggests that the imaginary part decreases with the charge implying that the real oscillation frequency will be increasing with charge whereas the damping decreases with it. It is in this sense that the behavior of the frequencies in both the Reissner-Nordstr\"{o}m and charged Gauss-Bonnet background is same. Now let us comment on the Gauss-Bonnet coupling $\alpha\to 0$ limit. It has been shown in \cite{konoplyaGB} and \cite{sayan} that in the $\alpha\to 0$ limit, the quasinormal modes for scalar field perturbations and vector and tensorial perturbations of the uncharged Gauss Bonnet black hole yields Schwarzschild QN frequencies. This is very easy to understand since in the limit $\alpha\to 0$, the Gauss Bonnet metric looks like $(1-2M/r^{d-3}+4\alpha M^2/r^{2d-4}+\cdots)$. The third term in the above metric is actually $\mathcal{O}(\alpha^2)$ term and hence for very small values of $\alpha$ the quasinormal frequencies for Gauss-Bonnet black hole goes to the Schwarzschild values. The same thing is expected here because in the $\alpha\to 0$ limit, the charged Gauss-Bonnet metric looks like $(1-2M/r^{d-3}+\tilde{Q}^2/r^{2d-6}+\mathcal{O}(\alpha^2)+\cdots)$, where $\tilde{Q}=Q/\sqrt{2\pi(d-2)(d-3)}$. We have checked for different small values of $\alpha ~(\alpha=0.0001, 0.0002,\cdots, 0.0005)$, keeping the charge fixed that the quasinormal frequencies for Reissner N\"{o}rdstrom black holes in different dimensions are indeed produced. 
 
\begin{figure}[here]
\centering
\subfigure[Plot of Real $\omega$ vs $\alpha$ for charged Gauss-Bonnet black hole]{\rotatebox{0}{\epsfxsize=5.5cm\epsfbox{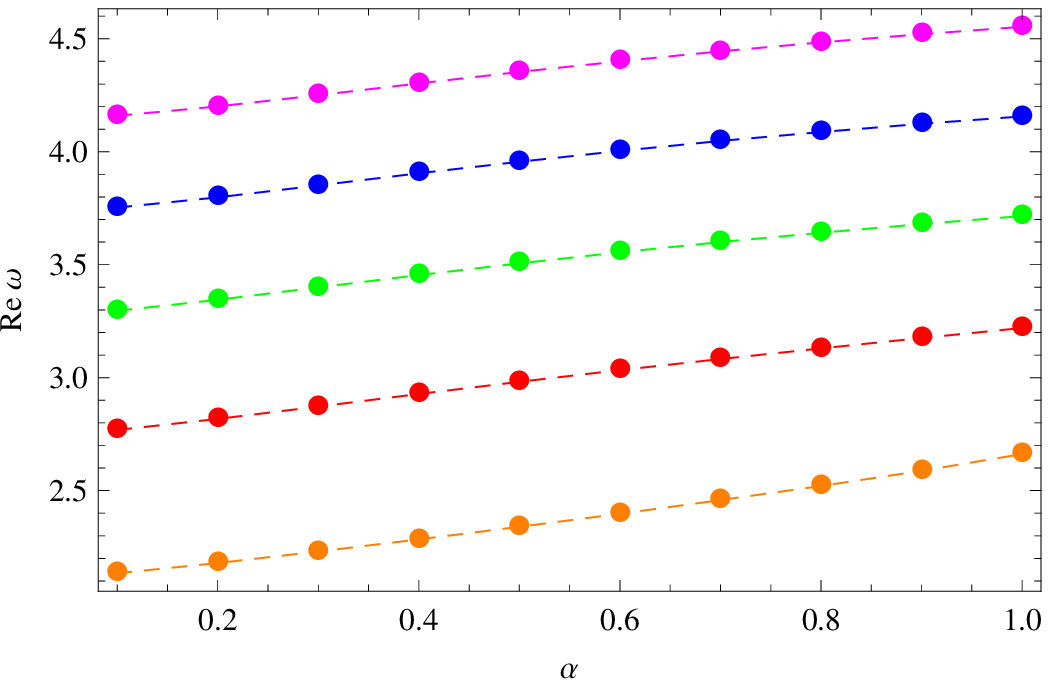}}}
%\hspace{0.1cm}
%\caption[]{The potential for higher dimensional Reissner-Nordstr\"{o}m like black holes. The lower one is for $d=5$ and the topmost one is for $d=10$ for charge $Q=0.05$ and $l=2$}
\subfigure[Plot of Imaginary $\omega$ vs $\alpha$ for charged Gauss-Bonnet black hole]{\rotatebox{0}{\epsfxsize=5.5cm\epsfbox{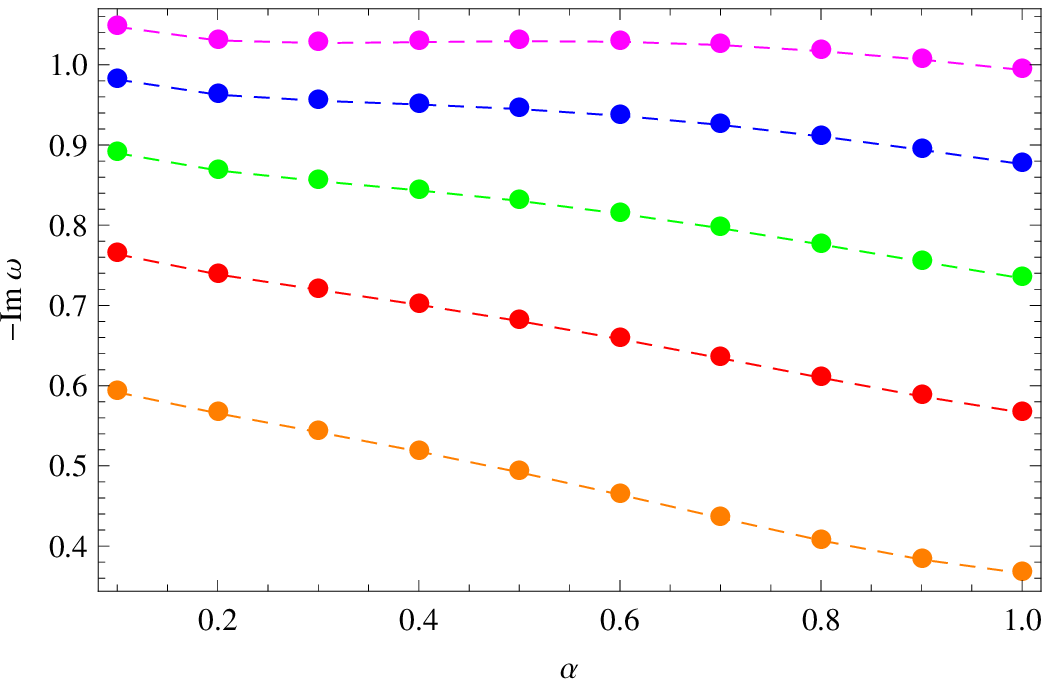}}}
%\end{center}
%\hspace{0.1cm}
\subfigure[Plot of Quality Factor vs $\alpha$ in $d=7$]{\rotatebox{0}{\epsfxsize=5.5cm\epsfbox{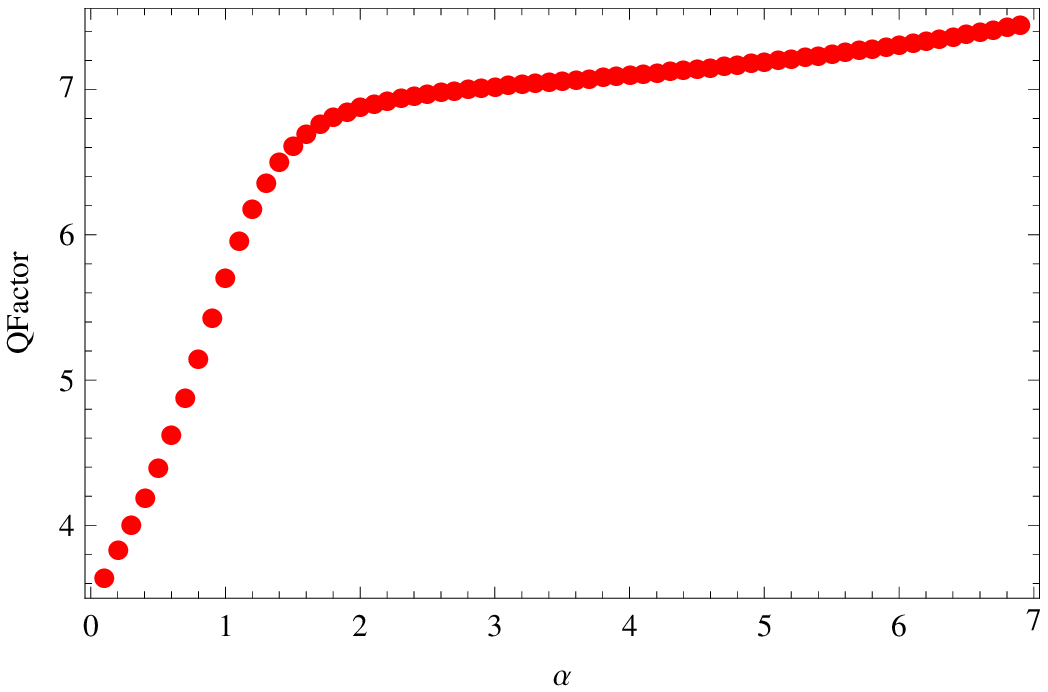}}}
\caption{The variation of (a) real, (b) imaginary parts of frequency and (c) quality factor with Gauss-Bonnet coupling $\alpha$ for a fixed value of charge $Q=0.05$ and multipole number $l=1$. Dimension of the space time increases for (a) and (b) from bottom to top, i.e. the lowest plot in each graph corresponds to $d=6$ and top one corresponds to $d=10$}
\label{alphGBQ}
\end{figure}

Having discussed the behaviour of the real and imaginary parts of the frequencies with the charge of the Gauss Bonnet black hole, let us now concentrate on the variation of the frequencies with the parameter $\alpha$, which is the Gauss-Bonnet coupling. The variation of the real and imaginary parts of the frequency with the Gauss-Bonnet coupling is plotted in Fig. (\ref{alphGBQ}). As can be seen from the figures, the real part of the frequency increases with the increase of Gauss-Bonnet coupling while the negative imaginary part decreases with $\alpha$. It may be noticed that the variation of the imaginary part in higher dimensions is comparatively slower than the lower dimensions. For example, in $d=8, ~9,~ 10$, the negative imaginary part first starts decreasing with $\alpha$, then remains unchanged for certain consecutive values and again starts decreasing with $\alpha$.

If one defines the quality factor as $\hat{Q}\sim\vert\frac{Re(\omega)}{Im(\omega)}\vert$, then its variation is plotted in fig (\ref{alphGBQ}c). This shows that in spite of the behaviour of the real and imaginary parts of the frequency stated above, the quality factor increases as one increase the value of $\alpha$. It is also interesting to note that though $\hat{Q}$ increases faster for small values of $\alpha$, it becomes somewhat saturated as $\alpha$ is gradually increased, which is similar to the behaviour of the quality factor observed for uncharged Gauss-Bonnet black hole on the brane also \cite{zhidGB}.    

\section{QNM in large multipole number limit}

Let us now focus on the large angular momentum (multipole number) limit, i.e. now $\kappa\to\infty$. Our aim in this section is to simplify the otherwise complicated expression for the frequency given by Eqn. (\ref{freq}) and get an analytic expression for $\omega$. For this, we will consider the large multipole number limit, where one can see from the expression of the potential (\ref{potl22}), the most dominant term of the potential in $\kappa\to\infty$ limit will be
\bea
V_1\vert_{\kappa\to\infty}\sim f\frac{\kappa^2}{r^2}.
\eea
Hence this approximation simplifies the potential a lot. Next, we will consider the WKB formula only upto the first order, i.e. now we will write Eqn. (\ref{freq}) as
\bea
\omega^2\sim V_0-i\left(n+\frac{1}{2}\right)(-2V_0^{\prime\prime})^{1/2}+\cdots \label{1ord},
\eea
where $V_0$ is the maxima of the potential $V_1$ that we are working with. For finding the maxima of the potential we solve the equation $\frac{dV_1}{dr}=0$ and get
\bea
r_m\vert_{\kappa\to\infty}\sim \left(\frac{2d-6}{d-3}\right)^{\frac{1}{d-3}}\frac{r_+r_-}{(r_+^{d-3}+r_-^{d-3})^{1/d-3}}
\eea
The form of the potential with this maxima of $r$ is
\bea
V_1\vert_{\kappa\to\infty}&=&\left(2^{\frac{1}{d-3}} r_- r_+
   \left(r_-^{d-3}+r_+^{d-3}\right){}^{\frac{1}{3-d}}\right){}^{
   4-2 d} \nonumber\\
&&\left(\left(2^{\frac{1}{d-3}} r_- r_+
   \left(r_-^{d-3}+r_+^{d-3}\right){}^{\frac{1}{3-d}}\right){}^{
   d-3}-r_-^{d-3}\right)\nonumber\\
&& \left(\left(2^{\frac{1}{d-3}} r_- r_+
   \left(r_-^{d-3}+r_+^{d-3}\right){}^{\frac{1}{3-d}}\right){}^{
   d-3}-r_+^{d-3}\right)
\eea
Next one uses this to find out an expression for the frequency
\bea
\omega^2&=&[2^{\frac{1}{d-3}} r_- r_+
   (r_-^{d-3}+r_+^{d-3}){}^{\frac{1}{3-d}}]{}^{
   4-2 d}\times [(2^{\frac{1}{d-3}} r_- r_+
   (r_-^{d-3}+r_+^{d-3}){}^{\frac{1}{3-d}}){}^{
   d-3}-r_-^{d-3}]\times\nonumber\\
&& [(2^{\frac{1}{d-3}} r_- r_+
   (r_-^{d-3}+r_+^{d-3}){}^{\frac{1}{3-d}}){}^{
   d-3}-r_+^{d-3}]\nonumber\\
&&-\frac{1}{4} i \left(n+\frac{1}{2}\right)\times\nonumber\\
&&\left(\frac{1}{r_-^2
   r_+^2}\right) [-2^{\frac{4-d}{3-d}}
   (r_-^{d-3}+r_+^{d-3}){}^{-\frac{2}{d-3}}
   (2^{\frac{1}{d-3}} r_- r_+
   (r_-^{d-3}+r_+^{d-3}){}^{\frac{1}{3-d}}){}^{
   -2 d} (-(d-1) d (2^{\frac{1}{d-3}} r_- r_+\nonumber\\
&& (r_-^{d-3}+r_+^{d-3}){}^{\frac{1}{3-d}}){}^d
   (r_+^3 r_-^d+r_+^d r_-^3)
   (r_-^{d-3}+r_+^{d-3}){}^{\frac{3}{d-3}}+3
   2^{\frac{d-6}{d-3}}\nonumber\\
&& (2^{\frac{1}{d-3}} r_- r_+
   (r_-^{d-3}+r_+^{d-3}){}^{\frac{1}{3-d}}){}^{
   2 d}
   (r_-^{d-3}+r_+^{d-3}){}^{\frac{6}{d-3}}+2^{\frac{d
   }{d-3}} (d (2 d-7)+6) r_-^{d+3} r_+^{d+3})]{}^{\frac{1}{2}}
\eea

Which indeed is very complicated, though we have simplified our potential and used the first order WKB formula. We can note from the above form that the QN frequencies indeed depend on the mass and charge of the black hole via $r_+$ and $r_-$ (since $r_+$ and $r_-$ are determined in terms of $\mu$ and $\theta$). One can check by putting $d=4$ in the above formula that the result for 4 dimensional Reissner Nordstr\"{o}m black hole is indeed obtained.

\section{Conclusion}

In this paper, we have discussed the Dirac quasinormal frequencies of charged black holes in higher dimensions. We have investigated two different backgrounds, namely the higher dimensional Reissner-Nordstr\"{o}m and the charged Gauss-Bonnet black holes. Though the field of studying QNMs is being saturated day by day, a comparative study of QNMs in different scenarios might be interesting. 

One of the main ideas of \cite{split} was to study fermion quasinormal modes in purely higher dimensional Schwarzschild background within the framework of split fermion models where the quarks and the leptons are forced to live on separate branes in order to keep proton stability. The present work can also be interpreted as studying split fermion quasinormal modes for higher dimensional charged black holes. Within this model, we found that the real part of the QN frequencies increases with the increase of the charge in both the backgrounds discussed in this paper, while the imaginary part decreases with the charge for both. The behavior of the frequencies are studied with the increases in space time dimensions. We found that the real part of the frequency increases as we increase the space time dimensions while the damping also increases. A comparative study of the quasinormal modes in the higher dimensional RN and charged Gauss-Bonnet black hole suggests that the variation of the frequencies with charge is much more rapid in the Reissner-Nordstr\"{o}m background than the charged Gauss-Bonnet background. Quantitatively we found that the real and imaginary parts of the frequencies in the split fermion models are larger than their brane localized partners. One can also get back the QN frequencies of the black holes arising out of pure Einstein theory of gravity from the values of the QN frequencies of charged Gauss-Bonnet background in the Gauss-Bonnet coupling $\alpha\to 0$ limit. We have also studied the behaviour of the real and imaginary parts of the frequencies with the Gauss-Bonnet coupling $\alpha$ and found that the real part increases with $\alpha$, while the negative imaginary part decreases. For higher dimensions this variation in the imaginary part is slower and it can be seen that for some consecutive values of $\alpha$, the change in the imaginary part is very small and then if we vary $\alpha$ again, the negative imaginary part starts decreasing again. 

It would be interesting now to calculate the black hole absorption cross section for bulk RN and Gauss Bonnet fermions in higher dimensions and a comparative study of late time fall off would also be interesting in these backgrounds. As it was found that the massive fields might change the low lying modes of the QN spectrum, the study of massive Dirac perturbations in these background will also be an important aspect.  

\section{Acknowledgment}

The author wishes to thank Prof. Kumar S. Gupta for numerous discussions and suggestions during the work. He also wishes to thank Prof. Palash B. Pal and Pulak Ranjan Giri for many useful discussions during the earlier stages of the work. Finally he wishes to thank Subhajit Karmakar for numerous help regarding programming with Mathematica.

\section{Appendix A: Dirac Equation in spherically symmetric higher dimensional black
  hole background}

In this section we will study the Dirac equation in higher dimensions in a static spherically symmetric black hole background. Essentially this section contains a brief review of the works done in \cite{split}.

Let us start with a $d$ dimensional spherically symmetric metric of
the form
\bea
ds^2=-f(r)dt^2+f^{-1}(r)dr^2+r^2d\bar\Omega_{d-2}^2, \label{mtrc}
\eea
where $d\bar\Omega_{d-2}^2$ is the metric for $(d-2)$
sphere. Following \cite{split,gibbons,gibbons1}, let us think
of a conformal transformation under which the metric and the
determinant of the metric behaves as 
\bea
g_{\mu\nu}\to\tilde g_{\mu\nu}&=&\Omega^2g_{\mu\nu}\nonumber \\ 
g^{\mu\nu}\to\tilde g^{\mu\nu}&=&\Omega^{-2}g^{\mu\nu}\nonumber\\
\mathrm{det}~|g_{\mu\nu}|&=&\Omega^{-2d}\mathrm{det}~|\tilde g_{\mu\nu}|\nonumber\\
\sqrt{-\tilde{g}}&=&\Omega^{d}\sqrt{-g}
\eea
where $\Omega$ is the conformal factor. Now, under such a conformal
transformation the spinor field behaves as
$\psi\to\tilde\psi=\Omega^n\psi$, where $n$ can be determined by
claiming that the Dirac Lagrangian remains invariant under such a
conformal transformation. Using the fact that
$\{\gamma^{\mu},\gamma^{\nu}\}=2g^{\mu\nu}$ and
$\tilde\gamma^{\mu}=\Omega^{-1}\gamma^{\mu}$ along with the massless
Dirac Lagrangian one can get $n=-(d-1)/2$. Thus
\bea
\psi\to\tilde\psi&=&\Omega^{\frac{(1-d)}{2}}\psi\nonumber\\
\gamma^{\mu}\nabla_{\mu}\psi\to\tilde\gamma^{\mu}\tilde\nabla_{\mu}\tilde\psi&=&\Omega^{-\frac{(d+1)}{2}}\gamma^{\mu}\nabla_{\mu}\psi
\eea
Following \cite{split}, we choose $\Omega^2=\frac{1}{r^2}$, then
$\tilde\psi=r^{(d-1)/2}\psi$. The conformal metric $d\tilde s^2$ then
has a completely separated $t-r$ part and the $(d-2)$-sphere
part. The massless Dirac equation then can be written as 
\bea
\tilde\gamma^{\mu}\tilde\nabla_{\mu}\tilde\psi=0 \label{dirac}
\eea
where
$\tn_{\mu}=\tilde\partial_{\mu}-\frac{i}{4}\tilde\eta_{ac}\tilde\omega^c_{b\mu}\tilde\sigma^{ab}$,
and $\tilde\omega^c_{b\mu}=\tilde e^c_{\nu}\tilde \partial_{\mu}\tilde
e^\nu_b+\tilde e^c_{\nu}\tilde e^\sigma_{b}
\tilde\Gamma^{\nu}_{\sigma\mu}$,
$\tilde\sigma^{ab}=\frac{i}{2}[\tg^a,\tg^b]$.
The problem now is to find out the $\tg$ matrices for higher dimensional space time. In \cite{zee}, a complete discussion about this problem was given for flat space time. Following their notation and using \cite{split} we write the separated Dirac equation as 
\bea
\left[\left(\frac{r}{\sqrt{f}}(-i\sigma^3)\tn_t+r\sqrt{f}\sigma^2\tn_r\right)\otimes \mathbf{1}\right]\p + \left[-\sigma^1\otimes(\tg^i\tn_i)_{S_{d-2}}\right]\p=0
\eea
Where, $\sigma^i$ are the Pauli matrices
\bea
\sigma^1 =
\left(\begin{array}{cc}
0 & 1 \\
1 & 0 
\end{array}\right),~~
\sigma^2 =
\left(\begin{array}{cc}
0 & -i \\
i &  0
\end{array}\right),~~
\sigma^3 =
\left(\begin{array}{cc}
1 & 0 \\
0 & -1
\end{array}\right)
\eea
Following \cite{higuchi}, one can write 
\bea
(\tg^i\tn_i)_{S_{d-2}}\t\chi_l^{\pm}=\pm
i\left(l+\frac{d-2}{2}\right)\t\chi_l^{\pm}, ~~~~l=0,~1,~2,\cdots
\eea
where $\t\chi_l^{\pm}$ are the eigenspinors for the
$(d-2)$-dimensional sphere. Again, following \cite{split}, one can
write $\p=\displaystyle\sum_{l}(\t\phi_l^+(r,t)\t\chi_l^++\t\phi_l^-(r,t)\t\chi_l^-)$ using the
orthogonality of the eigenspinors. The Dirac Equation can then be written as
\bea
\sigma^2 r
\sqrt{f}\left[\partial_r+\frac{r}{2\sqrt{f}}\frac{d}{dr}\left(\frac{\sqrt f}{r}\right)\right]\phi_l^+-i\sigma^1\left(l+\frac{d-2}{2}\right)\phi_l^+=i\sigma^3\left(\frac{r}{\sqrt{f}}\right)\partial_t\phi_l^+,
\eea
In writing the above
equation we have changed the notation by removing the tilde from the
expressions and also we are working here with the positive sign of the eigenspinors. It may be mentioned that one can also choose to work with the negative sign as well. Next, we use the ansatz 
\bea
\phi_l^+=\left(\frac{\sqrt f}{r}\right)^{-1/2}e^{-i\omega t}\left(\begin{array}{c}
iG(r)\\
F(r)
\end{array}
\right)
\eea
The Dirac equation then simplifies to 

\bea
f\frac{dG}{dr}-\frac{\sqrt{f}}{r}\left(l+\frac{d-2}{2}\right)G=\omega F,\\
f\frac{dF}{dr}+\frac{\sqrt{f}}{r}\left(l+\frac{d-2}{2}\right)F=-\omega G
\eea
By defining the tortoise coordinate as $dr_{\star}=dr/f(r)$ and introducing a function 
\bea
W=\frac{\sqrt{f}}{r}\kappa
\eea
where $\kappa=(l+(d-2)/2)$, one can finally arrive at
\bea
\left(-\frac{d^2}{dr_{\star}^2}+V_1\right)G=\omega^2G;~~~~\left(-\frac{d^2}{dr_{\star}^2}+V_2\right)F=\omega^2F,
\eea  
where
\bea
V_{1,2}=\pm \frac{dW}{dr_{\star}}+W^2.
\eea
The potentials $V_1$ and $V_2$ corresponding to Dirac particles and anti-particles are supersymmetric to each other and dervide from the same superpotential $W$. It has also been proved \cite{jing3} that the Dirac particles and anti-particles have the same QN spectra. In this paper we have used only $V_1$ to study the quasinormal modes.

\bibliographystyle{unsrt}

\begin{thebibliography}{99}
\bibitem{kk} Kokkotas K D and Schmidt B G 1999 {\it Living Rev. Rel.} {\bf 2} 2
\bibitem{nol} Nollert H-P 1999 {\it Class. Quantum Grav.} {\bf 16}
  R159
\bibitem{rg} Regge T and Wheeler J A 1957 {\it Phys. Rev.} {\bf 108} 1063
\bibitem{vish} Vishveshwara C V 1970 {\it Phys. Rev.} {\bf D1} 2870
\bibitem{sterio} Kokkotas K D and Stergioulas N {\it Preprint}
  gr-qc/0506083
\bibitem{fera} Ferrari V and Gualtieri L 2008 {\it Gen. Rel. Grav.}
  {\bf 40} 9458
\bibitem{card_thes} Cardoso V {\it Preprint} gr-qc/0404093
\bibitem{danny} Birmingham D, Sachs I and Solodukhin S N 2002
{\it Phys. Rev. Lett.} {\bf 88} 151301
\bibitem{danny1} Birmingham D, Sachs I and Solodukhin S N 2003 {\it
    Phys. Rev.} {\bf D67} 104026
\bibitem{horo} Horowitz G T and Hubeny V E 2000 {\it Phys. Rev.} {\bf
    D62} 024027;    
Konoplya R A 2002 {\it Phys. Rev.} {\bf
    D66} 044009
\bibitem{card1} Cardoso V and Lemos J P S 2001 {\it Phys. Rev.} {\bf
    D63} 124015
\bibitem{kono3} Konoplya R A 2003 {\it Phys. Rev.} {\bf
    D68} 124017; Konoplya R A and Zhidenko A 2004 {\it JHEP} {\bf
    06} 037; Zhidenko A 2004 {\it Class. Quant. Grav.} {\bf 21} 273; Cardoso V and Lemos J P S 2003 {\it Phys. Rev.} {\bf
    D67} 084020;  
Molina C 2003 {\it Phys. Rev.} {\bf
    D68} 064007; Lopez-Ortega A 2006 {\it Gen. Rel. Grav.} {\bf 38} 1565; Lopez-Ortega A 2006 {\it Gen. Rel. Grav.} {\bf 38} 743, Lopez-Ortega A 2007 {\it Gen. Rel. Grav.} {\bf 39} 1011, Lopez-Ortega A 2008 {\it Gen. Rel. Grav.} {\bf 40} 1379; Smirnov A A 2005 {\it Class. Quant. Grav} {\bf 22} 4021
\bibitem{perl} Perlmutter S {\it etal} 1997 {\it Astrophys. J.} {\bf
    483} 565
\bibitem{hod} Hod S 1998 {\it Phys. Rev. Lett.} {\bf 81} 4293 
; Motl L and Neitzke A 2003 {\it Adv. Theor. Math. Phys.}
  {\bf 7} 307;  
Das S and Shankaranarayanan S 2005 {\it Class. Quantum
  Grav.} {\bf 22}, L7; 
Nat\'{a}rio J  and Schiappa R 2004 {\it
    Adv. Theor. Math. Phys.} {\bf 8} 1001; 
Chakrabarti S K and Gupta K S 2006 {\it Int. Jour. Mod. Phys} {\bf A21} 3565;
 Daghigh R G and Kunstatter G 2005 {\it
    Class.Quant.Grav.} {\bf22} 4113; 
Kunstatter G 2003 {\it Phys.Rev.Lett.} {\bf 90}
  161301;
Maggiore M 2008 {\it Phys.Rev.Lett.} {\bf100}
  141301; Vagenas E C 2008 {\it JHEP} {\bf 0811} 073.
\bibitem{cho} Cho H T 2003 {\it Phys.Rev.} {\bf
    D68} 024003 
\bibitem{jing} Jing J 2005 {\it Phys. Rev} {\bf D71} 124006
\bibitem{lever} Lever E 1985 {\it Proc. Roy. Soc.} {\bf A402} 285 
\bibitem{hill} Majumdar B and Panchapakesan N 1989 {\it Phys. Rev}
  {\bf D40} 2568
\bibitem{khc} Castello-Branco K H C, Konoplya R A and Zhidenko A 2005
  {\it Phys. Rev.} {\bf D71} 047502
\bibitem{wu} Wu Y-J and Zhao Z 2004 {\it Phys. Rev} {\bf D69} 084015; 
Chang J-F and Shen Y-G 2007 {\it Int. J. Theor. Phys.}
  {\bf 46}, 1570;
Jing J 2005 {\it JHEP} {\bf 0512} 005; Chang J-F and Shen Y-G 2005 {\it Nucl. Phys.} {\bf
    B712} 347; Jing J L and Pan Q Y 2005 {\it Phys. Rev.} {\bf D71}
  124011
\bibitem{jing3} Jing J L 2004 {\it Phys. Rev.} {\bf D69} 084009
\bibitem{antoniadis1} Antoniadis, I 1990 {\it Phys. Lett.} {\bf B246} 377
\bibitem{led} Arkani-Hamed N, Dimopoulos S and Dvali G R 1998 {\it
    Phys. Lett.} {\bf B429} 263
\bibitem{antoniadis} Antoniadis I, Arkani-Hamed N, Dimopoulos S and Dvali G R 1998 {\it Phys. Lett.} {\bf B436} 257
\bibitem{kanti1} Kanti P and Konoplya R A 2006 {\it Phys. Rev.} {\bf
    D73} 044002
\bibitem{kanti2} Kanti P, Konoplya R A and Zhidenko A 2006 {\it Phys.
    Rev.} {\bf D74} 064008
\bibitem{zhidGB} Zhidenko A 2008 {\it Phys. Rev.} {\bf D78}, 024007 
\bibitem{split} Cho H T, Cornell A S, Doukas J and Naylor W 2007 {\it
    Phys. Rev} {\bf D75} 104005
\bibitem{gibbons} Das S R, Gibbons G W and Mathur S D 1997 {\it
    Phys. Rev. Lett.} {\bf 78} 417
\bibitem{gibbons1} Gibbons G W and Steif A R 1993 {\it Phys. Lett.}
  {\bf B314} 13  
\bibitem{tnln} Tangherlini F R 1963 {\it Nuovo Cimento} {\bf 77} 636
\bibitem{mype} Myers R C and Perry M J 1986 {\it Annals Phys.} {\bf 172} 304
\bibitem{zwei} Zwiebach B 1985 {\it Phys. Lett} {\bf B156}, 315 
\bibitem{deser} Boulware D G and Deser S 1985 {\it Phys. Rev. Lett.} {\bf
  55}, 2656 
\bibitem{wheeler} Wheeler J T 1986 {\it Nucl. Phys.}  {\bf B268}, 737 
\bibitem{wheeler1} Wheeler J T 1986 {\it Nucl. Phys.} {\bf B273}, 732 
\bibitem{wiltsh} Wiltshire D L 1988 {\it Phys. Rev.} {\bf D38}, 2445
\bibitem{moura} Moura F and Schiappa R 2007 {\it Class. Quant. Grav.} {\bf 24}, 361
\bibitem{chandra} Chandrasekhar S and Detweiler S 1975 {\it Proc. Roy.
    Soc.(London)} {\bf A344}, 441 
\bibitem{ferrari} Ferrari V and Mashhoon B 1984 {\it Phys. Rev.} {\bf
    D30}, 295 
\bibitem{will} Schutz B and Will C M 1988 {\it Astrophys. J.} {\bf
    291} L33 
\bibitem{will2} Iyer S and Will C M 1985 {\it Phys. Rev} {\bf D35}
  3621
\bibitem{iyer} Iyer S 1987 {\it Phys. Rev.} {\bf D35}, 3632 
\bibitem{kon2} Konoplya R A 2003 {\it Phys. Rev.} {\bf D68}, 024018
\bibitem{Andersson} Andersson N 1992 {\it Proc. R. Soc. (London)} {\bf
    A439}, 47 
\bibitem{and1} Andersson N and Linnaeus S 1992 {\it Phys. Rev.} {\bf
    D46}, 4179 
\bibitem{leaver} Leaver E W 1985 {\it Proc. R. Soc. (London)} {\bf
    A402}, 285 
\bibitem{konoplyaGB} Konoplya R 2005 {\it Phys. Rev.} {\bf D71} 024038 
\bibitem{konabdalla} Abdalla E, Konoplya R A and Molina C 2005 {\it Phys. Rev.} {\bf D72} 084006
\bibitem{sayan} Chakrabarti S K 2007 {\it Gen. Rel. Grav.} {\bf 39} 567 
\bibitem{konGBz} Konoplya R and Zhidenko A 2008 {\it Phys. Rev.} {\bf D77} 104004 
\bibitem{cardosoyousi} Cardoso V, Lemos J P S and Yoshida S 2004
  {\it Phys. Rev.} {\bf D69}, 044004
\bibitem{zee} Wilczek F and Zee A 1982 {\it Phys. Rev.}{\bf D25} 553
\bibitem{higuchi}Camporesi R and Higuchi A 1996 {\it J. Geom. Phys.}
  {\bf 20} 
\end{thebibliography}

\end{document}